\documentclass[useAMS,usenatbib]{mn2e}
\usepackage[cp1251]{inputenc} 
\usepackage{graphicx}
\usepackage{amssymb}
\usepackage[english]{babel}
\usepackage{morefloats}
\usepackage[pdftex]{hyperref}
\usepackage{ulem}
\usepackage[dvipsnames]{color}

\def\ltsima{$\; \buildrel < \over \sim \;$}
\def\simlt{\lower.5ex\hbox{\ltsima}}
\def\gtsima{$\; \buildrel > \over \sim \;$}
\def\simgt{\lower.5ex\hbox{\gtsima}}

\newcommand{\Xcounit}{cm\ensuremath{^{-2} } (K km s\ensuremath{^{-1} })\ensuremath{^{-1} }\,}
\newcommand{\kmps}{km~s\ensuremath{^{-1}}\,}
\newcommand{\Msun}{M\ensuremath{_\odot}\,}
\newcommand{\Oo}{\displaystyle}
\newcommand{\Lum}{K~km~s\ensuremath{^{-1} } pc\ensuremath{^{2 }}\,}

\title[GMCs scaling relations]{GMCs scaling relations: role of the cloud definition}

\author[Khoperskov et al]{ S.A. Khoperskov$^{1,2,3}$\thanks{sergey.khoperskov@unimi.it}, E.O. Vasiliev$^{4,5}$, D.A. Ladeyschikov$^{6}$,  \and A.M. Sobolev$^{6}$, A.V. Khoperskov$^{7}$\\
$^1$Dipartimento di Fisica, Universit\`{a} degli Studi di Milano, via Celoria 16, I-20133 Milano, Italy \\
$^2$Institute of Astronomy, Russian Academy of Sciences, Pyatnitskaya st., 48, 119017 Moscow, Russia\\
$^3$Sternberg Astronomical Institute, Moscow M.V. Lomonosov State University, Universitetskij pr., 13, 119992 Moscow, Russia\\
$^4$Institute of Physics,Department of Physics, Southern Federal University, Sorge 5, Rostov on Don 344090, Russia \\
$^5$Special Astrophysical Observatory, RAS, Nizhnii Arkhyz, Karachaevo-Cherkesskaya Republic,  369167 Russia \\
$^6$Ural Federal University, 51 Lenin Str., Ekaterinburg 620000, Russia \\
$^7$Volgograd State University, Universitetsky pr., 100, 400062 Volgograd, Russia}

\pagerange{\pageref{firstpage}--\pageref{lastpage}} \pubyear{2015}

\def\LaTeX{L\kern-.36em\raise.3ex\hbox{a}\kern-.15em
    T\kern-.1667em\lower.7ex\hbox{E}\kern-.125emX}

\begin{document}

\label{firstpage}

\maketitle
\begin{abstract}
We investigate physical properties of molecular clouds in disc galaxies with different morphology: a galaxy without prominent structure, a spiral barred galaxy and a galaxy with flocculent structure. Our $N$-body/hydrodynamical simulations take into account non-equilibrium H$_2$ and CO chemical kinetics, self-gravity, star formation and feedback processes. For { the} simulated galaxies the scaling relations of giant molecular clouds or so called Larson's relations are studied for two types of a cloud definition { (or extraction methods)}: the first { one} is based on total column density position-position (PP) datasets and the second one is indicated by the CO~(1-0) line emission used position-position-velocity (PPV) data. We find that the cloud populations obtained by using both cloud extraction methods generally have similar physical parameters. Except that for the CO data  the mass spectrum of clouds has a tail with low-massive objects  $M\sim 10^3-10^4$~\Msun. { Varying column density threshold the power-law indices in the scaling relations  are significantly changed.}
In contrast, the relations are invariant to CO brightness temperature threshold. { Finally, we find that the} mass spectra of clouds for the PPV data are almost insensitive to the galactic morphology, whereas the spectra for the PP data demonstrate significant variations.
\end{abstract}


\section{Introduction}

Molecular gas in galaxies is mostly concentrated in cold clouds with masses $\simeq 10^{4-5}$\Msun which are usually 
called giant molecular clouds (GMCs). Their evolution is important for understanding transition of gaseous 
component into the stellar one. Indeed, galactic star formation generally occurs in the dense medium of GMCs. \cite{1981MNRAS.194..809L}
initially introduced three empirical scaling relations for the nearby molecular clouds in the Milky Way (MW). These relations reflects the general view on the GMCs properties and have the following sense:

$\bullet$ cloud size - line-of-sight velocity dispersion relation, $\sigma_{\rm v} \propto R^{\rm \beta_1}$, is 
the first { one}, it argues that the cloud structure is supported by the internal turbulence;

$\bullet$ cloud virial mass - luminosity { in CO lines} relation, $M_{\rm vir} \propto  L^{\rm \beta_2}_{\rm CO}$, is 
the second { one}, it shows that GMCs are structures in the virial equilibrium;

$\bullet$ { luminosity in CO lines (sometimes cloud mass used)} -- size, $L_{\rm CO} \propto  R^{\rm \beta_3}$, is the third { one}, it { claims}
that the mean cloud surface density $\Sigma_0$ is likely to be constant { if} $\beta_3 \approx 2$.

Despite a long way of the scaling relations investigation, a complete theoretical explanation for the origin of the relations has not
been offered yet. Based on the CO observations of molecular clouds in the Galactic disc it has been found that GMCs have approximately
constant surface density $\sim 170$~\Msun~pc$^{-2}$ and the state of the clouds is really close to the virial equilibrium 
\citep{1987ApJ...319..730S}. \citet{0004-637X-723-1-492} have found tight power-law correlation with index $2.36\pm 0.04$ between radii and 
masses of the Galactic molecular clouds. The virial parameter of the derived clouds is mostly below $1$ with the mean value $0.46$, so
that clouds are strongly self-gravitating. Using $^{12}$CO data \citet{2009ApJ...699.1092H} re-examined the scaling relations for 
the Galactic clouds under assumption of
constant CO-to-H$_2$ conversion factor within a cloud. This leads to lower median mass surface value, which is 42~\Msun~pc$^{-2}$.
Note that the clouds found in this study are mostly unbound, that is in contradiction to the previous studies. Thus, the observational data demonstrates significant scatter in the physical state of GMCs even in the Milky Way.

For { molecular clouds in} both dwarf and giant disc galaxies \citet{2008ApJ...686..948B} have found scaling relations similar to that for the Milky Way clouds.
They have concluded that GMCs identified on its CO emission are the unique class of objects that exhibits a remarkably uniform set of properties from galaxy to galaxy. Meanwhile more recent comparison of GMCs in nearby galaxies by~\citet{2013ApJ...779...46H} let to figure out that { the} GMCs properties { (mass, radius, velocity dispersion)} are not robust towards to the external conditions: { clouds are}  smaller and fainter in less dense regions, { i.e. inside} low-mass galaxies and the outer regions of the Galaxy, compared to molecular structures in denser environment, e.g. in the inner part of the Galaxy and other spirals like M~51~\citep{2014ApJ...784....3C} and M~33~\citep{2003ApJS..149..343E,2010ApJ...725.1159B}.

Certainly, the scaling relations can reflect some universality in both physical conditions inside clouds and interaction of clouds with the ambient medium. Giant molecular clouds properties and evolution are governed by the interplay between self-gravity, magnetic field and feedback processes from stars born inside clouds. In many theoretical studies there has been attempted to understand how various feedback processes influence on the properties of GMCs~\citep{2008ApJ...684..978S,2011ApJ...730...11T,2012MNRAS.421.3488H,2014MNRAS.442.3407B}. For instance, \citet{2011MNRAS.413.2935D} have traced the evolution of individual clouds in detail and found that cloud-cloud collisions and stellar feedback can regulate the internal velocity dispersion and lead to formation of unbound GMCs. Contrary to the previous study \citet{2009ApJ...700..358T} suggested that molecular clouds are gravitationally bound because of the low collisional rate of clouds relatively to its orbital time scale. Thus, the internal turbulent energy can keep molecular clouds in the virial equilibrium. Several simulations of turbulence in GMCs \citep{2013MNRAS.436.1836R,2013MNRAS.436.3247K} have justified that self-gravity plays an important role in the cloud structure, but doesn't strongly affect on the 'velocity dispersion - size' relation. 

Using high resolution simulations \citet{2013ApJ...776...23B} analyzed the physical properties of clouds{ whose} number density { is above} $100$~cm$^{-3}$. They found that the slopes of the 'velocity dispersion - size' and 'mass-size' relations { appear to be} much steeper than the observational ones. On the other hand, \citet{2009ApJ...700..358T} got a good agreement between the mass, radius, velocity dispersion of GMCs with those observed in the Galaxy. Such contrary conclusions are explained by not only differences in simulations, e.g. taking into account star formation and other processes, but also variety in samples of clouds { caused by using different} methods of cloud extraction. Moreover, \citet{2014MNRAS.439..936F} found a significant effect of galactic environment on { the} cloud properties in the dynamical model of M~83. At first they established that the 'mass-size' relation has bimodal distribution, and at second, GMCs tend to be less gravitationally bound in denser environment, { i.e.}  spiral { arms} or bar, than in rarefied ones, e.g. { inside} disc. 

In { numerical} simulations a cloud is usually defined as an object whose gas density (column or { volume}) is higher than a given threshold.  Such { an object} can consist of { several} dense molecular cloud lets surrounded by diffuse { intercloud molecular and/or} atomic gas. { In addition} there are several other methods for cloud definition based on dust extinction, molecular or/and atomic column density or CO intensity. For each method it is interesting to find the scaling relations and compare it with the empirical ones established by Larson. That probably allows us to understand better what ISM structures are responsible for appearance of these relations.

The matching of the observed and simulated GMCs properties is not obvious because of different approaches used for cloud definition. In general, this problem has no unique solution both in observations, { because} in observations the border of a cloud can depend on a chosen signal-to-noise limit. { In numerical simulations} there are two commonly used methods { for} cloud extraction. The first { one} is based on total column density position-position (henceforth PP) data sets and the second { one} is indicated by the CO { line emission used position-position-velocity (PPV) data}. The latter is utilized in CLUMPFIND~\citep{1994ApJ...428..693W}, CPROPS~\citep{2006PASP..118..590R} packages. 

In this paper we consider { the} physical properties ({ namely,} mass, radius, surface density, velocity dispersion, luminosity etc.) of clouds for two methods of cloud extraction based on PP and PPV datasets. In our simulations we study the scaling relations or so-called Larson's laws for three MW-size galaxies with different morphology. The paper is organized as follows. Section~\ref{seq::models} contains the description of our numerical model. Section~\ref{seq::CD} describes methods of cloud definition. In Section~\ref{sec::phys_param} we present the statistical analysis of { the} physical properties of molecular clouds. Section~\ref{seq::scaling_relations} describes the scaling relations, the dependence of power-law indices of the relations on threshold value and mass spectra of GMCs for the simulated galaxies. In Section~\ref{sec:summary} we summarize our key results.

\section{Model}\label{seq::models}

To simulate{ the} galaxy evolution we use { our} code based on the unsplit TVD MUSCL (Total Variation Diminishing Multi Upstream Scheme for Conservation Laws) scheme for gas dynamics and{ the} $N$-body method for stellar component dynamics. In { gas} dynamical approach we reach the second order { in time}  and the third order{ in space} using the minmod limiter. For the Riemann problem solution we adopt the HLLC (Harten-Lax-van Leer-Contact) method. More details about gas dynamic part of our code can be found in the paper \citet{2014JPhCS.510a2011K}. Stellar dynamics is calculated using the second order flip-flop integrator. For the total stellar-gaseous gravitational field calculation we use the TreeCode approach.

For all models presented { here} we use a uniform grid with $4096\times4096\times512$ cells for { gas} dynamics and set a computational domain
$40 \times 40 \times 3$~kpc with spatial resolution 6~pc.{ The} initial number of{ stellar} particles is equal to $0.5 \times 10^6$, { during the simulation} it reaches  $2\times10^6$ depending on star formation activity.

\begin{table*}
\caption[]{
Initial parameters adopted in the simulations. Here the following notations are assumed:
$M_{\rm h}$ is the mass of dark matter halo within 12~kpc sphere,
$a_{\rm h}$ is the halo scale length,
$\sigma_r(0)$ is the central radial velocity dispersion,
$\sigma_{\rm z}/\sigma_{\rm r}$ is the ratio of the vertical velocity dispersion to the radial one,
$\Sigma_{\rm g0}$ is the central surface density of gaseous disc,
$h_g$ is the radial scale length of the gaseous disc,
$N^{\rm th}_{\rm tot}$ is the number of clouds extracted with the $N^{\rm th}_{\rm CDN}$~threshold (CDN is a cloud definition when the total column density within the cloud exceeds the $N^{\rm th}_{\rm tot}$ threshold and the cloud is delineated by corresponding level of the total column density),
$N_{\rm CF}$ is the number of clouds extracted using CLUMPFIND~(henceforth we use CF abbreviation for shortness). 
The following parameters are the same for considered models:
$\Sigma_{*0} = 835$ \Msun~pc$^{-2}$ is the central stellar surface density,
$h_* = 3$~kpc is the radial scale length of stellar disc, and
$h_{\sigma}* = 2h_*$ is the radial velocity scale length, $\Sigma_{\rm g0} = 10$~\Msun~pc$^{-2}$ is the central gas surface density.}
\begin{center}
\begin{tabular}{lcccccc|ccccccccc}
\hline
Model (Morphology) & \multicolumn{2}{c}{Halo} & \multicolumn{2}{c}{Bulge} & \multicolumn{2}{c}{Stellar disc} &  \multicolumn{4}{c}{Cloud definition}  \\
          & $M_{\rm h}$ & $a_{\rm h}$ & $M_{\rm b}$ & $b_{\rm h}$ & $\sigma_{\rm r(0)}$ &$\sigma_{\rm z}/\sigma_{\rm r}$ &    $N_{\rm CDN} $ & $N_{\rm CF}$ \\
          & $10^{10}$~\Msun & kpc & $10^{10}$~\Msun & kpc & km s$^{-1}$ &   & &   & \\
\hline
B (No structure) & 8.8   & 3.857 & -      & -          & 75 	&  0.5     & 1095 &  1150 \\
F (Milky Way like) & 8.8   & 1.1      & 0.7  & 0.153 & 100&  0.7    & 1065 &  1203 \\
H (Flocculent)& 8.25 & 1.1     & -      & - 		  & 50  	& 0.45    & 1012 &  1111 \\
\hline
\end{tabular}\label{tab::tabular1}
\end{center}
\end{table*}

\subsection{Chemical kinetics and gas thermodynamics}

Usually the emission in CO lines is the major source of the information about the GMCs
\citep{2001ApJ...547..792D,2008ApJ...686..948B,2009AJ....137.4670L}, and the intensity in CO lines is used to restore the mass of molecular hydrogen through $X_{\rm CO}$ factor \citep{1975ApJ...202...50D,2013ARA&A..51..207B}. Then, we are interested in a reasonable CO chemical network that on one hand gives fine CO molecule evolution and on the other requires adequate computational resources. Rather detailed networks include more than $20$ chemical species involved in several hundreds of reactions~\cite[e.g.][]{2000ApJ...534..809O, 2010MNRAS.404....2G}, which is computationally unacceptable for our purposes. Fortunately, \cite{2012MNRAS.421..116G}
found that the reduced network proposed by \cite{1999ApJ...524..923N} gives adequate results in comparison to the detailed chemical model, which consists of $218$ reactions amongst $32$ species \citep{2010MNRAS.404....2G}. So that here we exploit the model based on the network proposed by \cite{1999ApJ...524..923N}.

Based on our simple model for H$_2$ chemical kinetics~\citep{2013MNRAS.428.2311K} we expand the \citet{1999ApJ...524..923N} network by
several reactions needed for hydrogen ionization and recombination. For H$_2$ and CO photodissociation we use the approach described
by~\citet{1996ApJ...468..269D}. The CO photodissociation cross section is taken from \citet{2009A&A...503..323V}. In our radiation
transfer calculation described in Section~2.3 below we get ionizing flux at the surface of a computational cell. To calculate self-shielding
factors for CO and H$_2$ photodissociation rates and dust absorption factor for a given cell we use local number densities of gas and molecules, e.g. $f_{sh}^{\rm H_2} = n_{\rm H_2} L$, where $n_{\rm H_2}$ is H$_2$ number density in a given cell and $L$ is its physical size. The chemical network equations is solved by the CVODE package~\citep{Hindmarsh2005}.

We assume that a gas has solar metallicity with the abundances given in \citet{2005ASPC..336...25A}: $[{\rm C/H}] = 2.45 \times 10^{-4}, [{\rm O/H}] = 4.57\times 10^{-4}, [{\rm Si/H}] = 3.24\times 10^{-5}$. Dust depletion factors are equal to 0.72, 046 and 0.2 for C, O and Si, correspondingly. We suppose that silicon is singly ionized and oxygen stays neutral.

For cooling and heating processes we extend our previous model~\citep{2013MNRAS.428.2311K} by CO and OH cooling rates
\citep{1979ApJS...41..555H} and CI fine structure cooling rate~\citep{1989ApJ...342..306H}. The other cooling and heating rates
are presented in detail in Table~2~\cite[Appendix B in][]{2013MNRAS.428.2311K}. Here we simply provide a list of it: cooling due to recombination and collisional excitation and free-free emission of hydrogen~\citep{1992ApJS...78..341C}, molecular hydrogen
cooling \citep{1998A&A...335..403G}, cooling in the fine structure and metastable transitions of carbon, oxygen and silicon
\citep{1989ApJ...342..306H}, energy transfer in collisions with the dust particles \citep{2003ApJ...587..278W} and recombination cooling on the dust \citep{1994ApJ...427..822B}, photoelectric heating on the dust particles
\citep{1994ApJ...427..822B,2003ApJ...587..278W}, heating due to H$_2$ formation on the dust{ particles}, and the H$_2$ photodissociation
\citep{1979ApJS...41..555H} and the ionization heating by cosmic rays \citep{1978ApJ...222..881G}. In our simulations we achieve
gas temperature value as low as 10~K and number density as high as $5\times 10^3$~cm$^{-3}$.

\begin{figure*}
\includegraphics[width=0.49\hsize]{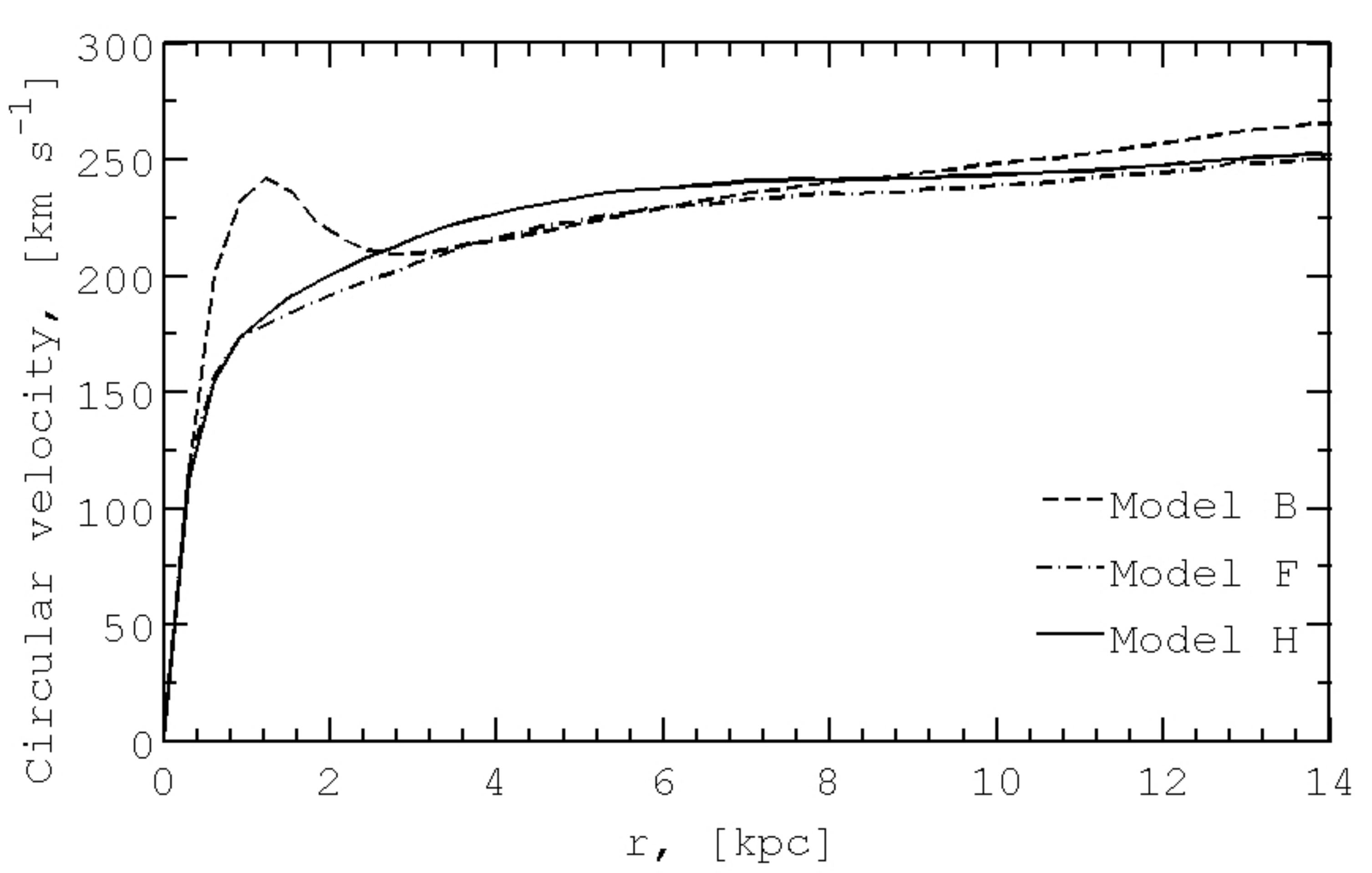}
\includegraphics[width=0.49\hsize]{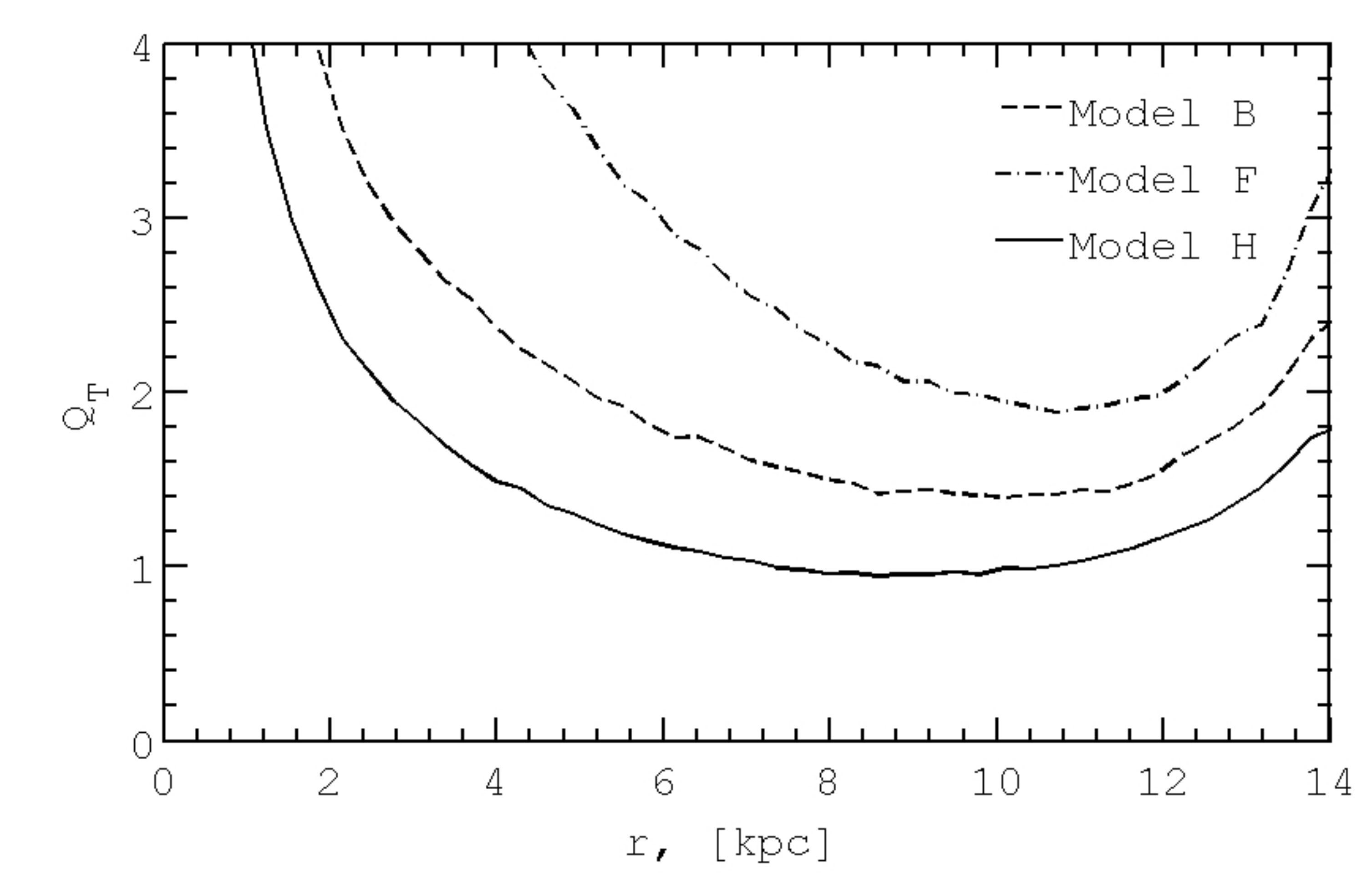}
\caption{Initial conditions for simulated stellar-gaseous discs. Circular velocity is shown in the left panel. Stability parameter $Q_T$ adopted for finite thickness two component disc and using the approximation by~\citet{2011MNRAS.416.1191R}is shown in the right panel. Physical parameters of galaxy models are presented in Table~\ref{tab::tabular1}.}\label{fig::initial_curves}
\end{figure*}

\subsection{Star formation and feedback}

In the star formation recipe adopted in our model mass, energy and momentum from the gaseous cells, where a star formation criterion is satisfied, are transited directly to newborn stellar particles. A star particle is formed in a grid cell, if the following criteria are fulfilled: (i) the gas density in the  cell should be higher than $100$~cm$^{-3}$ (such a high value prevents the formation of huge number of stellar particles), (ii) the total mass of gas in  surrounding cells exceeds the Jeans mass $M_{\rm cell}>M_{\rm J}$ (this help us to avoid the star formation in hot and warm medium, where some feedback processes occur), and we adopt the local star formation efficiency $0.01$. In star-forming cells the number density and temperature reach $n>200-500$~cm~$^{-3}$ and T\ltsima 50~K.

Feedback model includes several sources of thermal energy, namely stellar radiation, stellar winds from massive stars and SN explosions. The amount of injected energy connected to these processes is calculated for each stellar particle using the stellar evolution code~{\sc STARBURST99}~\citep{sb99}. We model supernova feedback only as thermal energy injection into a gas. We take into account mass loss by stellar particles due to SN explosions and stellar winds from both massive and low-mass stars.

\subsection{Radiation transfer}

To account molecule photodestruction we should know spatial structure of UV background in the galactic disc. Recent observations provide some evidences for significant radial and azimuthal variations of UV flux in the nearby galaxies~(Gil de Paz et al. 2007). No doubt that such variations are stipulated by local star formation. So that we need to include radiation feedback from stellar particles in our calculations.

Through our simulations the UV emission of each stellar particle is computed with the stellar evolution code {\sc STARBURST99}~\citep{sb99} assuming solar metallicity of stellar population. So that for each particle we know its luminosity evolution. After that we separate particles in two groups: young stellar particles (the age is smaller than $20$~Myr) and the other ones.
For definiteness we assume a uniform background field ten times lower than that in the Solar neighbourhood, $F_{b}=0.1$~Habing.Thus the UV background $F^{\rm UV}$ in a hydrodynamical cell with coordinates $\Oo {\bf {\rm r}_0}$ can be written as
\begin{equation}
\label{eq::UVequation}
 F^{\rm UV}({\bf {\rm r}_0}) = F_{\rm b} +  \sum_{i} F^{\rm old}_{i}({\rm r}_0) + \sum_{j} F^{\rm young}_{j}({\rm r}_0,{\rm r}_j)\,,
\end{equation}
$\sum_{i} F^{\rm
old}_{i}({\rm r}_0)$ is deposit from old stellar population (age $>20$~Myr), which plays a role only in a cell where the stellar
particle locates~($\Oo {\bf {\rm r}_0}$). The last term is UV flux from young stellar population -- the brightest stars. Their deposit is the most important in photodestruction of molecules in surrounding medium.

Due to  number of young stars is small at each time step, we can use the ray-tracing approach for each stellar particle. For $j$-th "young particle" we estimate the{ radius of} spherical shell (similar to the Stroemgren sphere), where the UV field value decreases down to 0.1~Habing:
\begin{equation}
\Oo R^{\rm d}_j = 0.1 \delta \sqrt{L^*_j/ (4\pi)}\,,
\end{equation}
where $L^*_j$ is luminosity of $j$-th stellar particle in Habing units and $\Oo \delta = \sqrt{(\delta x)^2+(\delta y)^2+(\delta z)^2}$ is effective cell size. For each shell we calculate the UV flux assuming the optical depth $\tau = 2  N / (10^{21} \rm{cm^{-2}})$,
where $N$ is the total column density of gas in cm$^{-2}$. So that we can get the distribution of the UV intensity in the entire galactic disc according to Eq.~\ref{eq::UVequation}.

\begin{figure*}
\includegraphics[width=0.49\hsize]{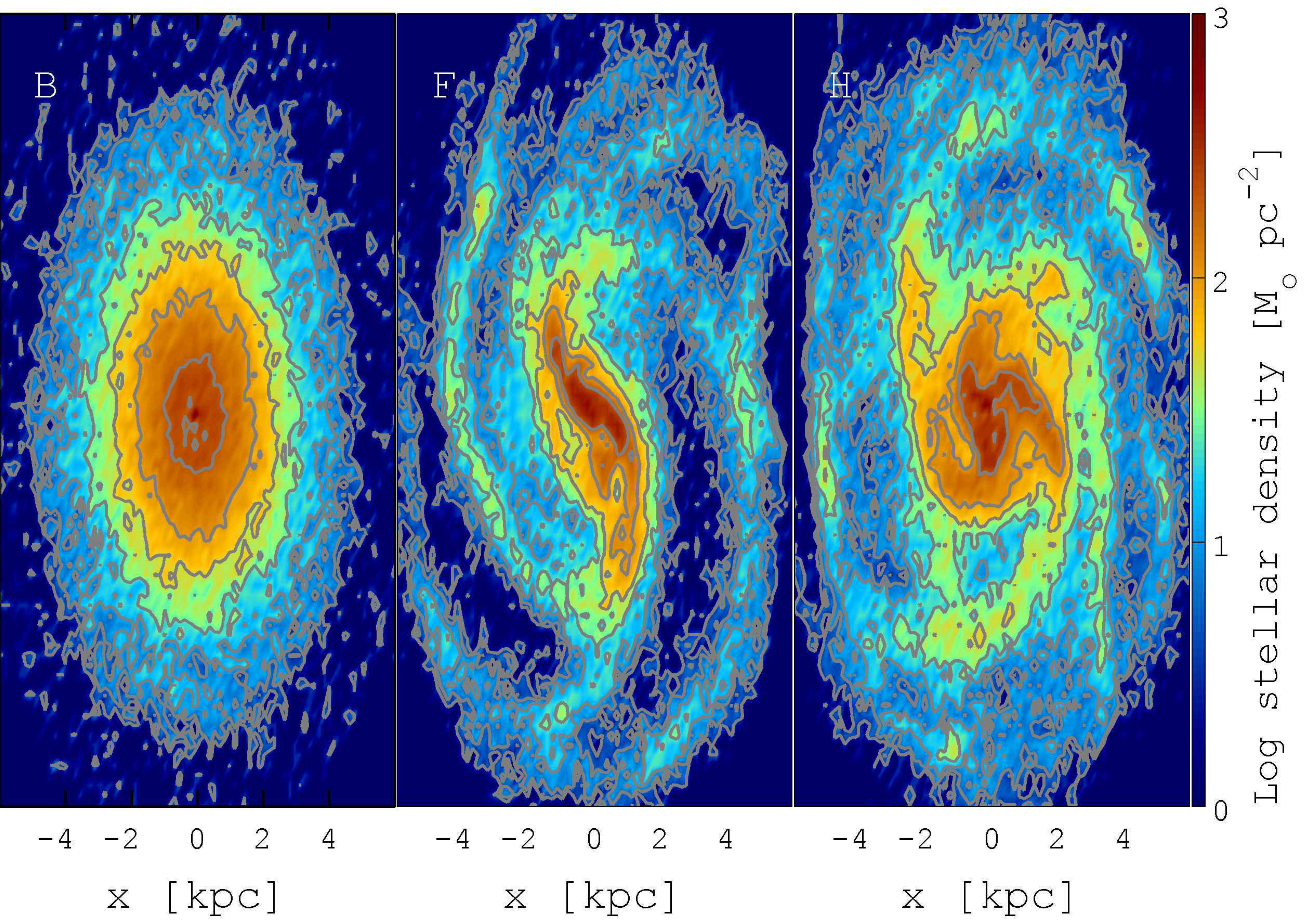}
\includegraphics[width=0.49\hsize]{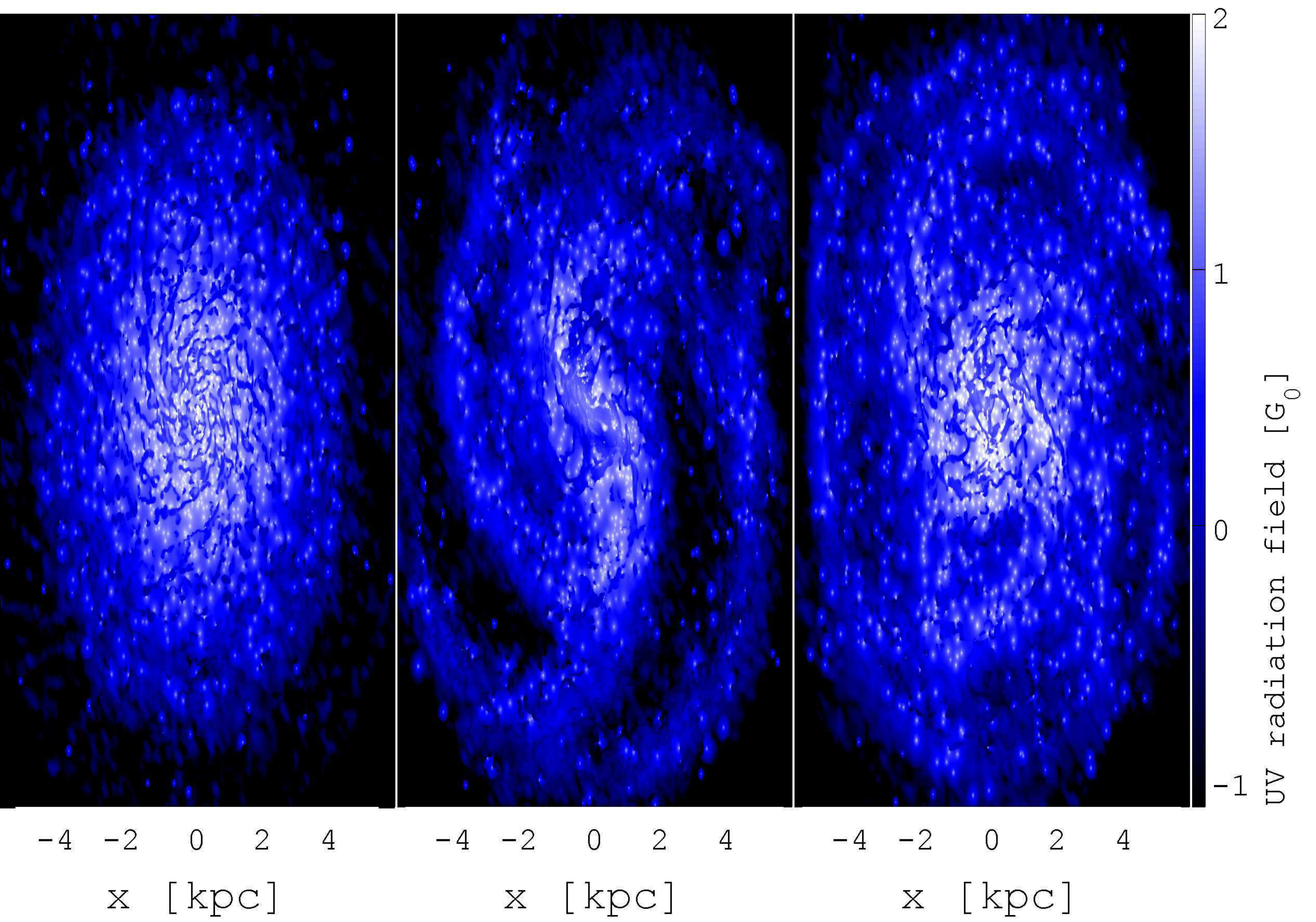}
\includegraphics[width=0.49\hsize]{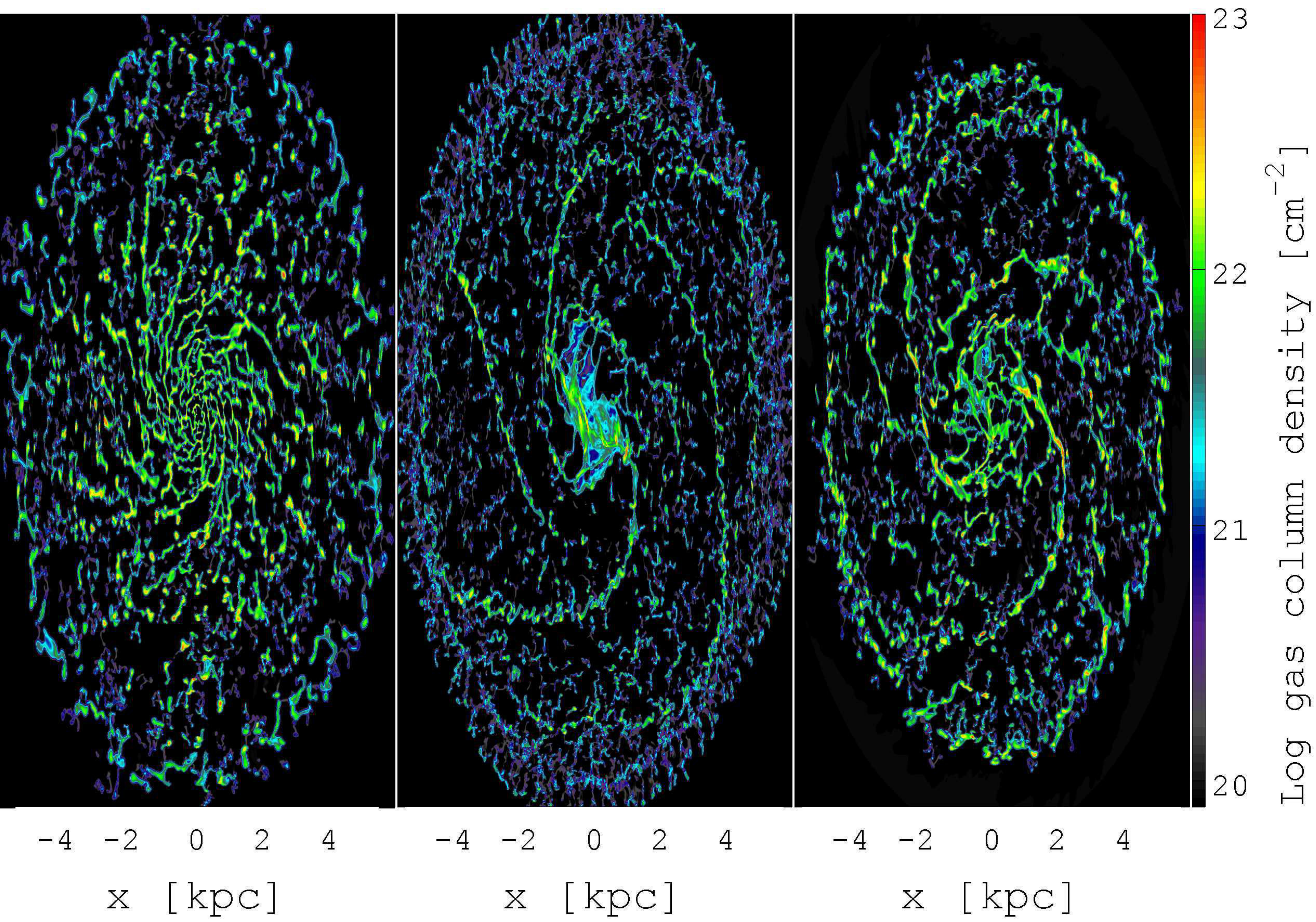}
\includegraphics[width=0.49\hsize]{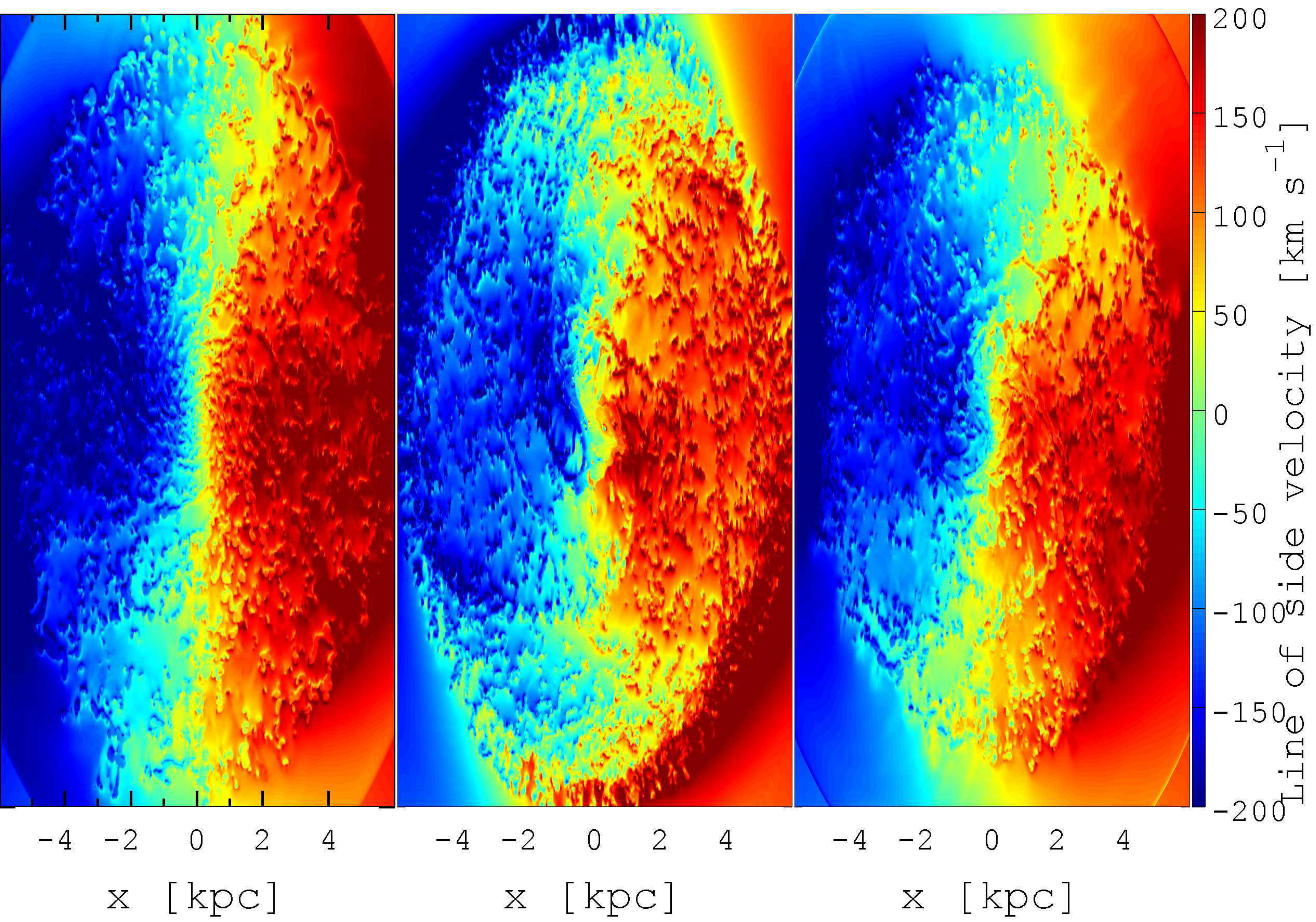}
\includegraphics[width=0.49\hsize]{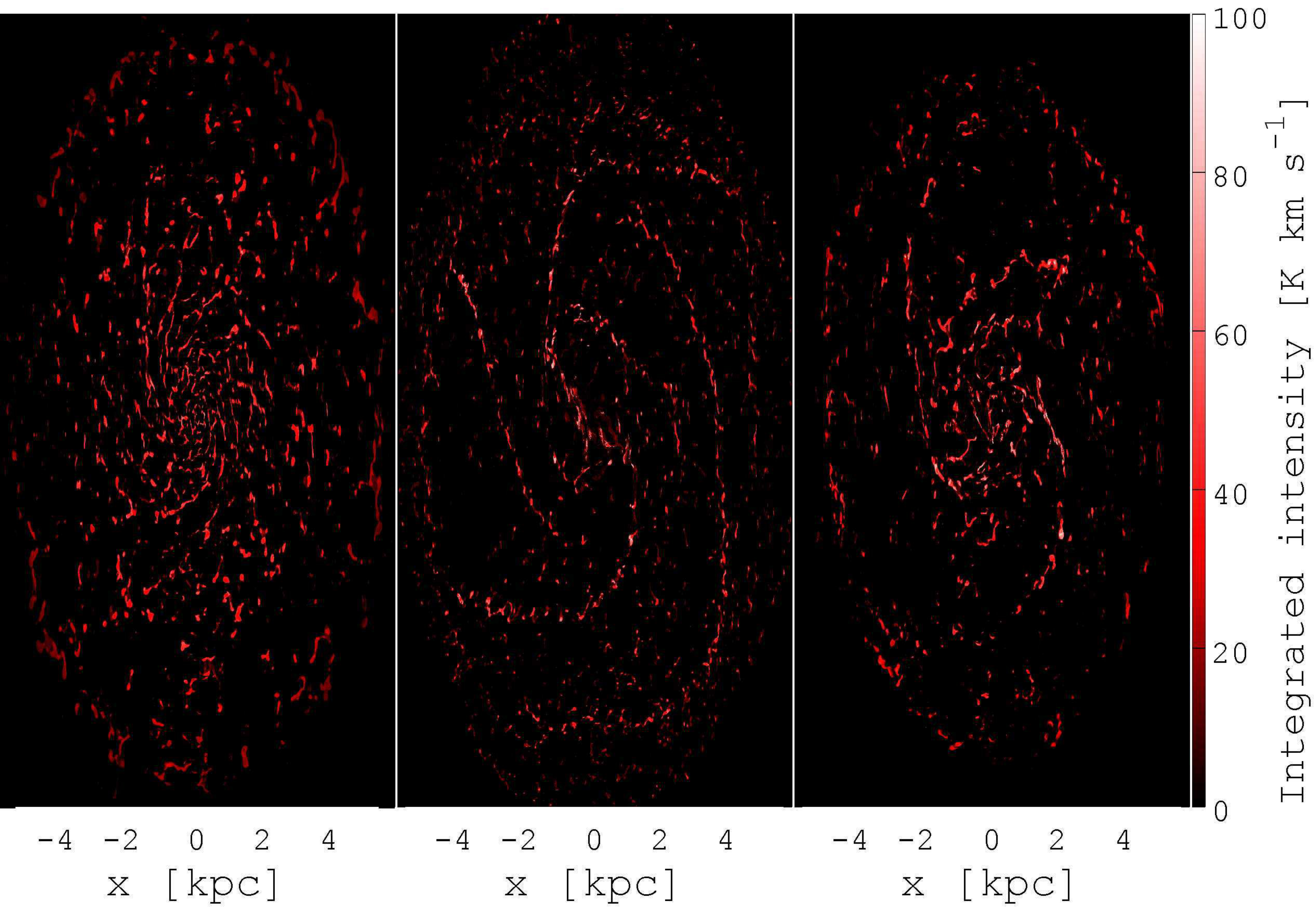}
\caption{
The projected maps (inclination $30^\circ$) of stellar surface density (top left group of panels), stellar UV radiation field (top right group of panels), total gas column density
(left group of panels in the middle row), radial (line-of-sight) velocity component (right group of panels in the middle row) and CO integrated intensity (bottom group of panels) at $t=500$~Myr for the following models of galaxies: no spiral structure or
model B (left map in all groups of panels), Milky Way-like or model F (central map in all groups of panels) and flocculent galaxy or model H (right map in all groups of panels). Initial parameters of the galactic models can be found in Table~\ref{tab::tabular1}.
}\label{fig::galaxies}
\end{figure*}

\subsection{Model of galaxies}

We start our simulations from the self-consistent radial and vertical equilibrium state of stellar-gaseous discs in the fixed
gravitational potential of dark matter halo. We assume both stellar and gaseous discs have exponential form, but with different
spatial scale lengths. Circular velocity of the gaseous disc embedded into the gravitational potential can be found as:
\begin{equation}
\Oo \frac{V^2_{\rm c}}{r} = - \left(\frac{\partial \Psi_{\rm halo}}{\partial r} + \frac{\partial \Psi_{\rm bulge}}{\partial r} + \frac{\partial \Psi_{\rm disc}}{\partial r} + \frac{\partial \Psi_{\rm gas}}{\partial r}\right)\,,
\end{equation}
where $\Psi_{\rm halo}$ is gravitational potential of dark matter halo, the halo is assumed to be steady isothermal
sphere,  $\Psi_{\rm bulge}$ is potential of bulge, $\Psi_{\rm disc}$ is potential of stellar disc and $\Psi_{\rm gas}$ is potential of gas. The parameters of{ the} gravitational potential can be found in  Table~\ref{tab::tabular1}.{ Fig.~\ref{fig::initial_curves} presents t}he radial dependence
of circular velocity for the galactic models considered here.

For stellar particle kinematics the asymmetric drift is taken in the form of Jeans approximation:
\begin{equation}\label{eq::jeans_eq}
\Oo V^2 = V^2_{\rm c} - \sigma^2_{\rm r} \left( 1 - \frac{\sigma^2_{\varphi}}{\sigma^2_{\rm r}} + \frac{r}{\Sigma_*} \frac{1}{\sigma^2_r}\frac{\partial (\Sigma_* \sigma^2_{\rm r})}{\partial r}\right)\,,
\end{equation}
where $\sigma_{\rm r}$, $\sigma_{\varphi}$ are radial and azimuthal velocity dispersions,{ respectively,}
$\Sigma_*$ is stellar surface density distribution.
The parameters of the potential and matter distributions can be found in Table~\ref{tab::tabular1}.
To compute an initial distribution of stars we solve Eq.~\ref{eq::jeans_eq} using the iterative procedure described in \cite{2003ARep...47..357K}.

Despite the parameters { of} the models{ of galaxies presented in Table~\ref{tab::tabular1}} are { close} to each other, the various stability conditions { allow} us to follow galaxies with the different morphology.
Initial stability criteria for two-component models (stellar-gaseous) are shown in the right panel of Fig.~\ref{fig::initial_curves}.
We compute three models of the stellar disc equilibria: gravitationally over-stable disc~(Model F, without prominent structure), highly unstable~(Model H, flocculent spirals morphology) and intermediate state disc~(Model B, MW-like morphology).{ The} initial stability parameter for two-component disc model accounting the finite disc thickness { effect}  is adopted in the form by~\cite{2011MNRAS.416.1191R}~(Fig.~\ref{fig::initial_curves}).

Figure~\ref{fig::galaxies} shows the maps of the stellar surface density, stellar UV radiation field, total gas column density and CO integrated intensity at $t=500$~Myr for the following models of galaxies inclined by $i=30^\circ$:
{ a galaxy without}spiral structure { (}model B),{ a} Milky Way type{ galaxy} (model F) and{ a} flocculent galaxy  (model H).{ The i}nitial parameters of the models are given in Table~\ref{tab::tabular1}. Note that the { spatially} averaged UV radiation field{ in all three models of galaxies} is significantly greater than a value of 0.1 Habing (see upper middle row in the Fig.~\ref{fig::galaxies}), so that our choice of the uniform background is reasonable. We have adopted inclination angle $i=30^\circ$  as a value which is enough to get significant line-of-sight velocity { scatter} while the structures in the { gaseous} disk are still rather well spatially distinguishable.

\section{Clouds definition}\label{seq::CD}

Prior to the calculation of physical parameters of clouds, we should define what the cloud is. In the most obvious approach for cloud definition
(CD) a cloud is an isolated gaseous clump with gas density~(column or { volume}) higher than a given level. This is the simplest criterion, but it doesn't reflect the chemical composition of a clump and we cannot say anything about  molecular content of such cloud. Moreover{ the} methods{ based on total gaseous column density} are not relevant to observable values, because using such methods some material, which is not associated with cloud itself and laid along the line of sight, can be regarded as a part of a cloud. This reveals in physical parameters of a cloud. So that we need a criterion based on the distribution of molecules in the ISM. Such criterion connects both chemical and extinction properties of a cloud and allows us to separate two phases of the cold interstellar medium: atomic and molecular gas.

Usually molecular clouds are studied by its emission in molecular lines~(e.g., for $^{12}$CO~\citep[see e.g.][]{1987ApJ...319..730S}, $^{13}$CO~\citep[see, e.g.][]{2009ApJ...699.1092H} and more recently in OH by~\citet{2015AJ....149..123A}). Since our model includes the H$_2$ and
CO molecule kinetics, we can use CD criteria based on both total gas column density and intensity in CO lines. This allows us to check the range of applicability for each CD criterion. So that the properties of a particular cloud { are expected to depend significantly} on the extraction criterion. It is also not clear how a choice of criterion influences on the statistical properties of the whole ensemble of molecular clouds.

Below we consider two approaches related to{ the} properties of a cloud measured in  observations. We define a cloud as a region inside that the total (molecular and atomic) hydrogen column density is higher than a given threshold $N^{\rm th}_{\rm tot}$~ (henceforth we use the abbreviation CDN for the method and corresponding indices). According to the CDN criterion we firstly find all local maxima (peaks) of  the gas column density in the plane of the galactic disc. After that around each local maximum we find cells with value higher  than a given threshold. In some cases these regions merge into larger ones and form a cloud with several local maxima. Usually  such coalescences take place in dense galactic structures, e.g. in spiral arms, bar or regions nearby the galactic center.{ Thus, o}ur approach  for finding clouds is a combination of the {\it 'contour method'} \citep{2014MNRAS.439..936F} and {\it 'peaks method'}~\citep{2009ApJ...700..358T}.

One of the widely used method for extracting structures from PPV data cubes is CLUMPFIND \citep{1994ApJ...428..693W}. This method is based on contouring data array at many different levels starting from the peak value and moving down to specific threshold. In the present work the CUPID implementation of{ the} CLUMPFIND{ algorithm} is used for { CO intensity} method of the clouds extraction~\citep{2007ASPC..376..425B}. Henceforth we use the abbreviation CF for { this} method and corresponding indices of variables.

We calculate CO brightness temperature in the form of the PPV data cubes using the method described in \cite{2012ApJ...758..127F}. In the calculations the spectral velocity resolution equals to $0.5$~\kmps, which is potentially enough { to resolve} structure of massive clouds.This spectral resolution is comparable to { that} one { reached} in the recent{ interferometric} observations~\citep[see e.g.,][]{2009ApJ...699.1153R}. Note that we discuss the dependence of power-law indices of the relations on velocity resolution{ value in the Section 5.4}.

Certainly, using two above-mentioned criteria we get two different population of clouds. Number and total mass of clouds are also different and depend on the value of column density { and} brightness temperature thresholds. In our analysis we take $N^{\rm th}_{\rm tot} = 1.9\times 10^{21}$~cm$^{-2}$ and  $T^{\rm th}_{\rm b} = 3$~K as fiducial threshold values, which provide us comparable numbers of clouds (around 1000) and similar total gaseous masses locked in clouds ($ M_{\rm t} \approx 2-4\times 10^{9}~\Msun$). Note that these values of $M_{\rm t}$ are close to the total mass of molecular clouds in the Milky Way~\citep{1997ApJ...476..166W}. The  numbers{ of clouds} and total gaseous masses { for the galactic models considered here} are given in Table~\ref{tab::tabular1}.

\begin{figure}
\includegraphics[width=1\hsize]{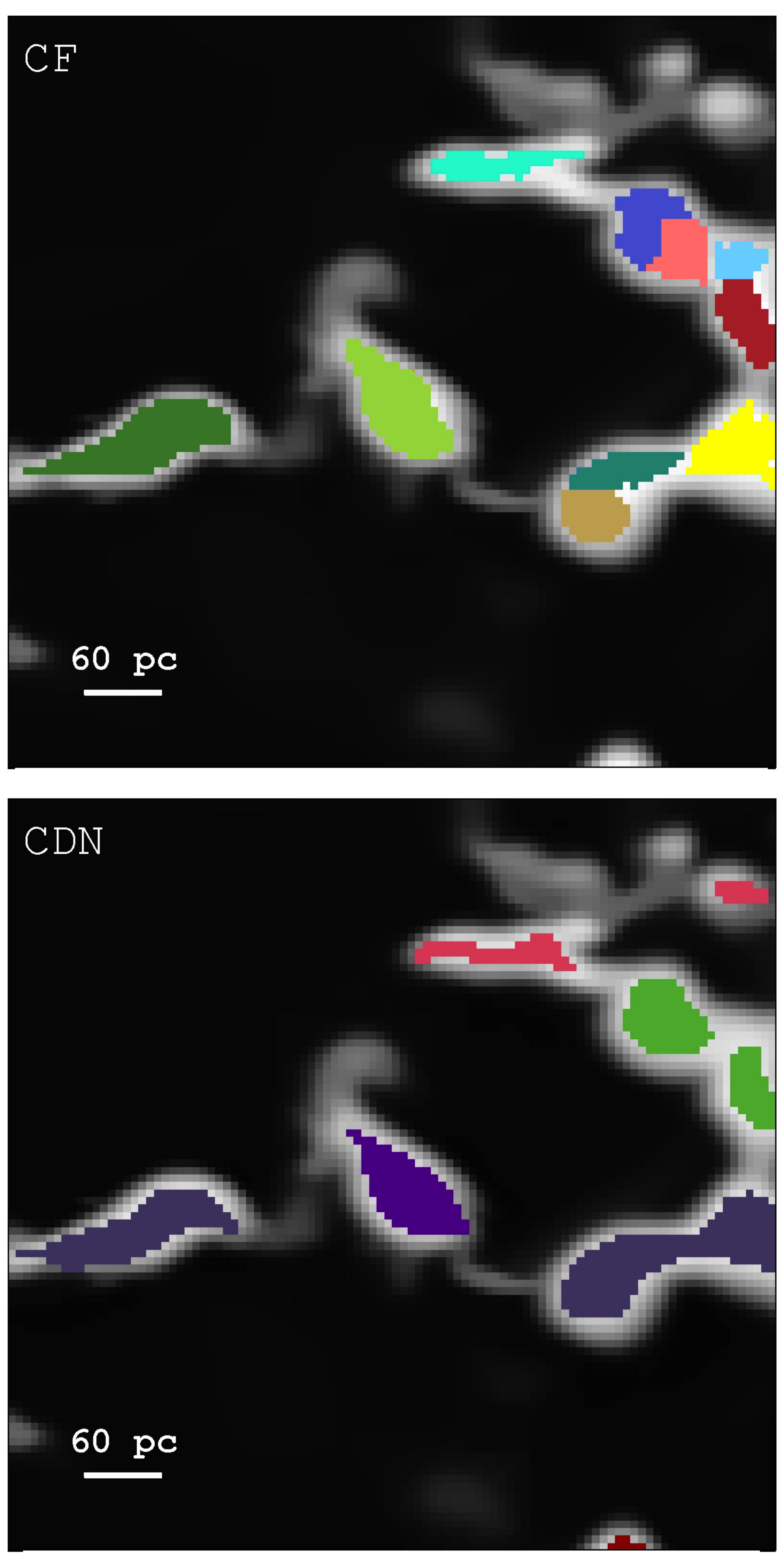}
\caption{
A region with the spatial distribution of extracted clouds in the model F. The extracted clouds are marked by color: top panel shows the distribution for CF (or CLUMPFIND) method, bottom panel demonstrates that for CDN approach. The background greyscale map is the gas surface density.}
\label{fig::clouds_clouds}
\end{figure}

For instance, a small region with the spatial distribution of extracted clouds in the {MW type galaxy} (model F) is shown in the Fig.~\ref{fig::clouds_clouds}. We mark the extracted clouds as coloured areas in contrast to the grey scale background of{ gaseous} column density { map}. It is clearly seen that spatial distributions and numbers of clouds are { remarkably distinct} for{ the} considered criteria. Prior to any quantitative analysis of the physical parameters we should notice two{ issues}. On one hand,  non-interacting clouds{ are appeared to} have similar shape for both extraction methods. { However}, in more dense environment clouds extracted by different methods look very unalikely.{ We suppose that t}his can be a result of dynamical effects related to cloud collisions and/or stellar feedback effects. On the other hand, it seems that very large clouds (or agglomerations) extracted by{ using the CDN method} have internal structure, which we can hardly resolve because of our spatial resolution{ is still not enough high}. Thus, { for the} CDN criterion large clouds and cloud chains (at least in the dense environment) { can be} extracted, while { using} the CF method such large structures are { splitted} into individual lumps with internal motions and other specific inhomogeneities.


\section{GMCs physical parameters}\label{sec::phys_param}

{ Cloud formation were studied numerically in detail by \cite{2006MNRAS.371.1663D,2008MNRAS.389.1097D,2008MNRAS.391..844D}. We mention that in our simulations  clouds {are result of }  self-gravity, thermal instability, cloud collisions and other processes occurred in the galactic disc. } {Here we briefly describe physical parameters of the cloud samples obtained in our analysis.}

On the one side, spiral arms stimulate GMCs formation due to gas falling into{ the gravitational} potential well of the arms. { Gravitational potential} of spiral structure induces collisions of clouds that in turn stimulates star formation. From the opposite side, supernovae explosions in star forming spiral arms can destroy clouds. { So one can conclude} that molecular clouds mostly form in spiral arms and are { probably} short-lived structures with lifetime \ltsima$ 10^7$~ yr~\citep{2009ApJ...699.1153R,2015arXiv150404528M}. However,  the existence of clouds in the inter-arm regions requires  longer  lifetime \citep{1979IAUS...84..277S,2009ApJ...700L.132K}. So that a question about lifetime of molecular clouds is still under debates \cite[see e.g.][]{2013MNRAS.432..653D,2014Ap&SS.353..595Z}.

\begin{figure*}
\includegraphics[width=0.49\hsize]{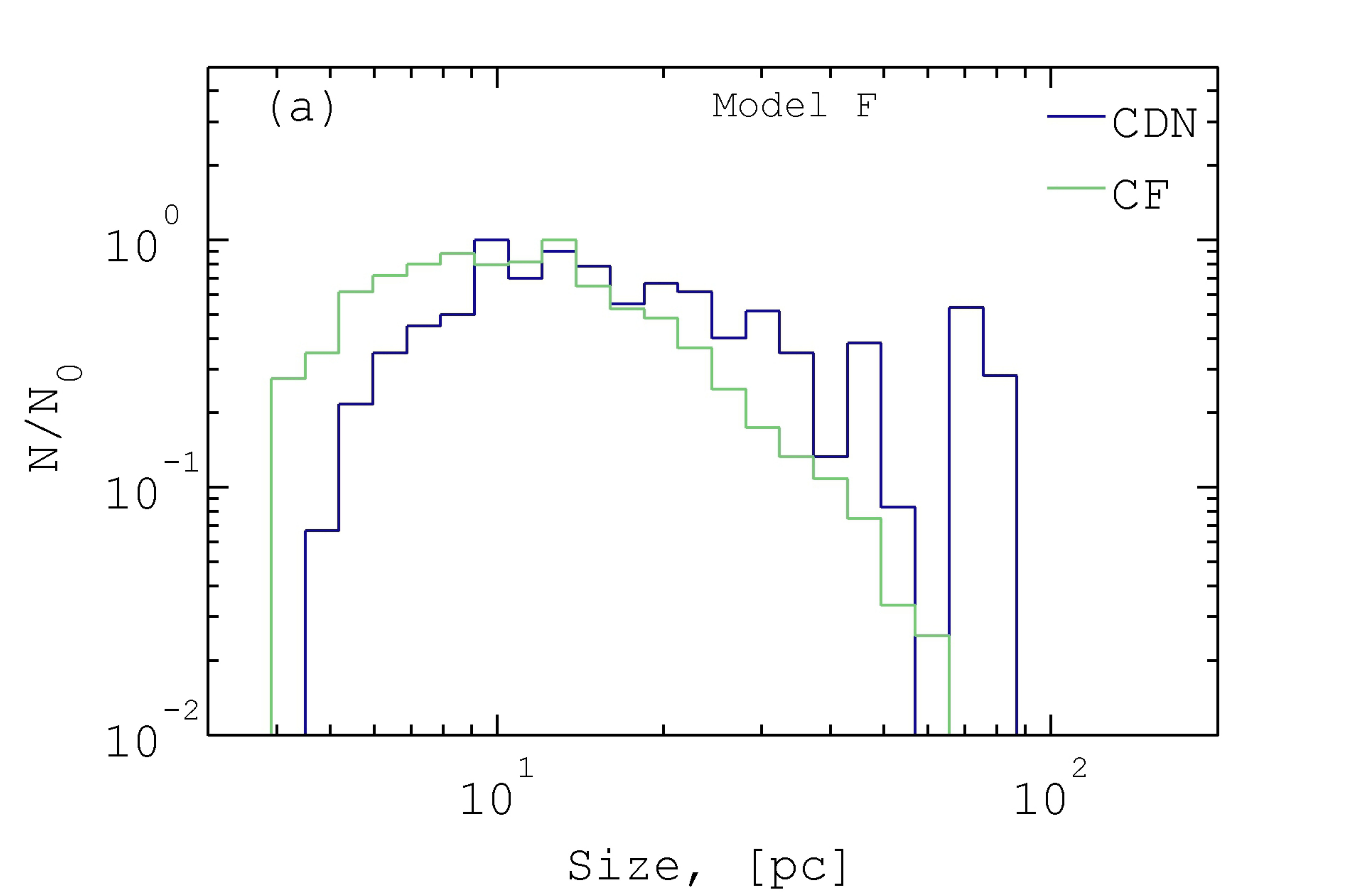}
\includegraphics[width=0.49\hsize]{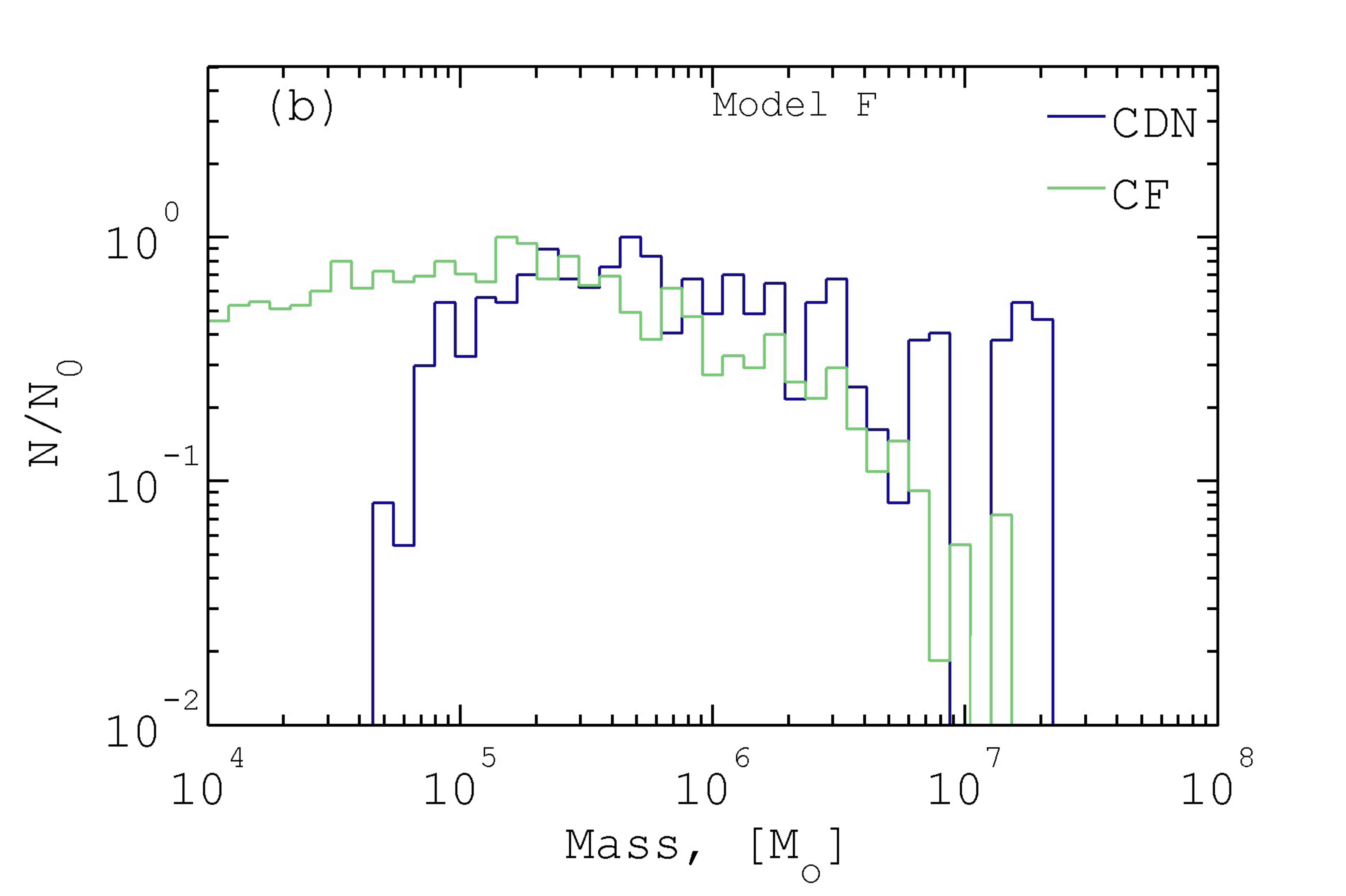}
\includegraphics[width=0.49\hsize]{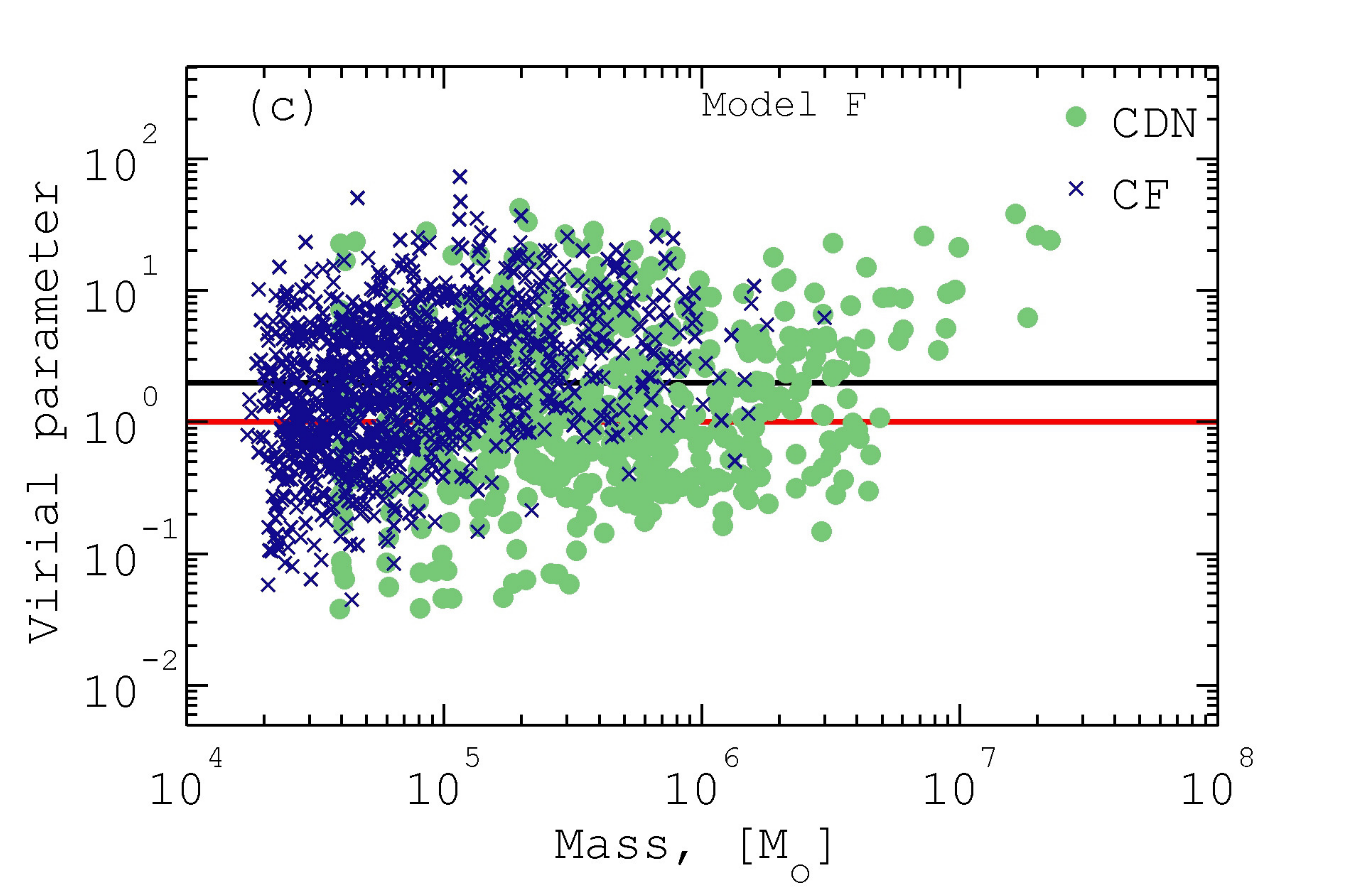}
\includegraphics[width=0.49\hsize]{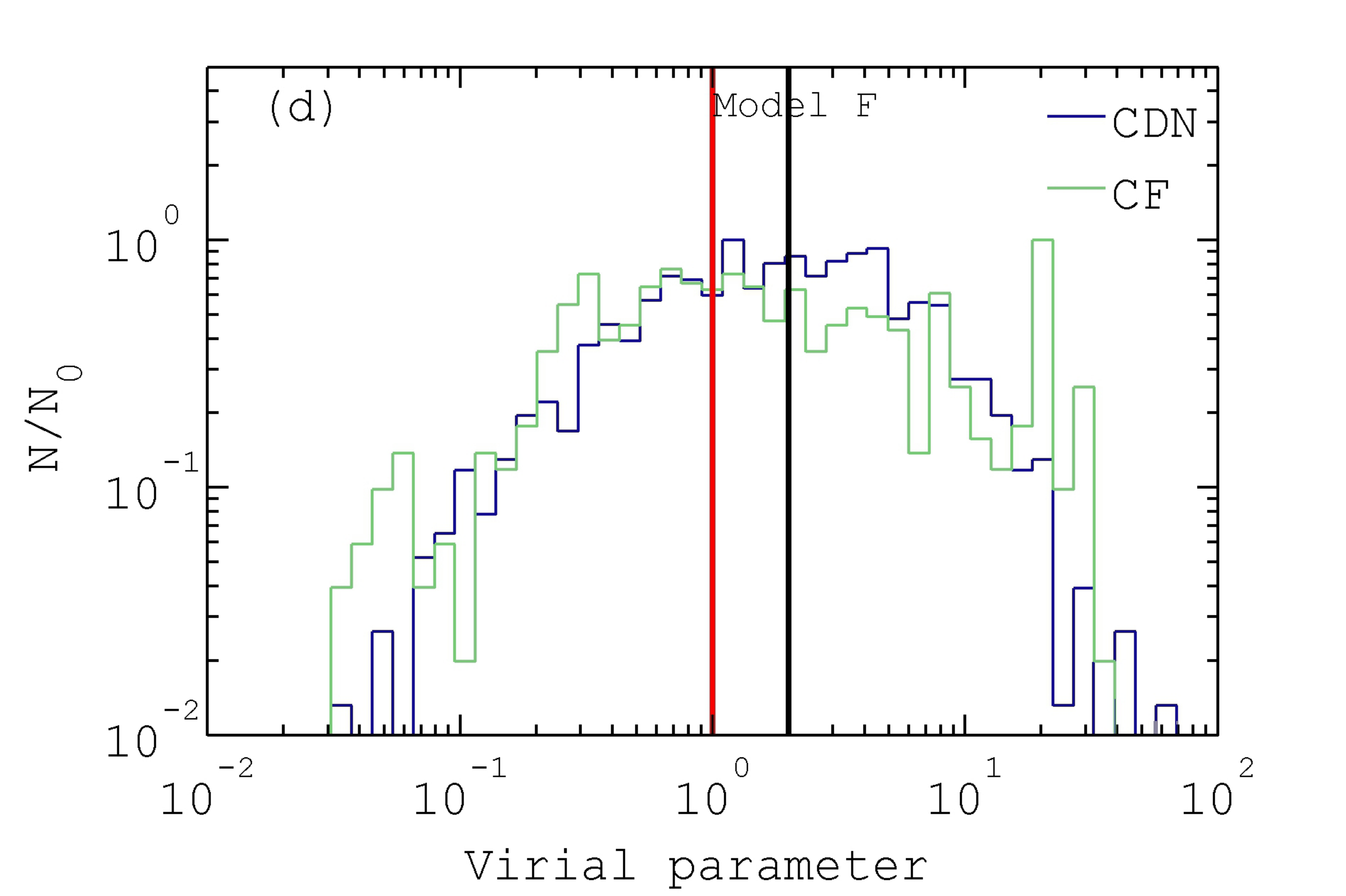}
\includegraphics[width=0.49\hsize]{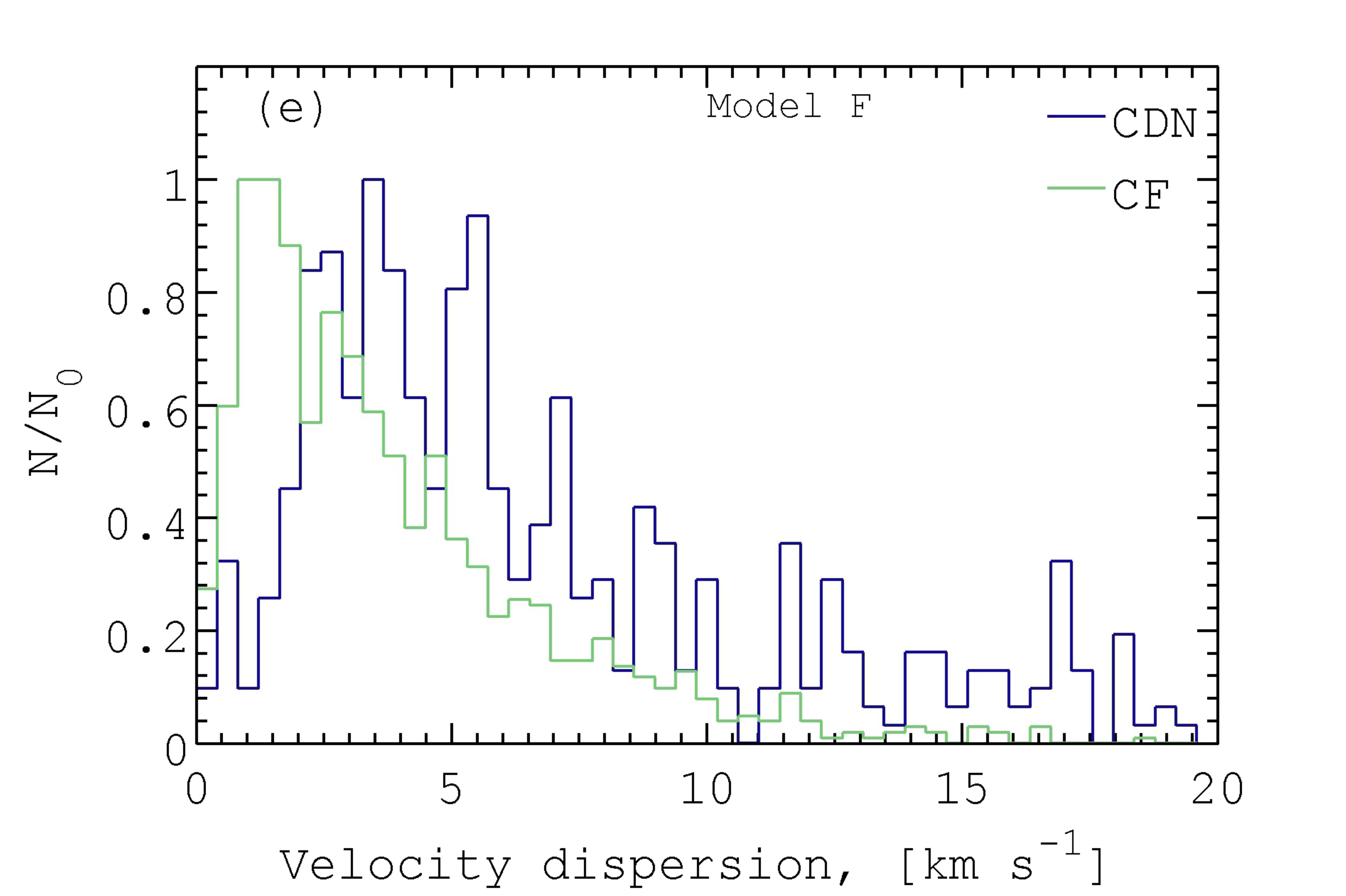}
\includegraphics[width=0.49\hsize]{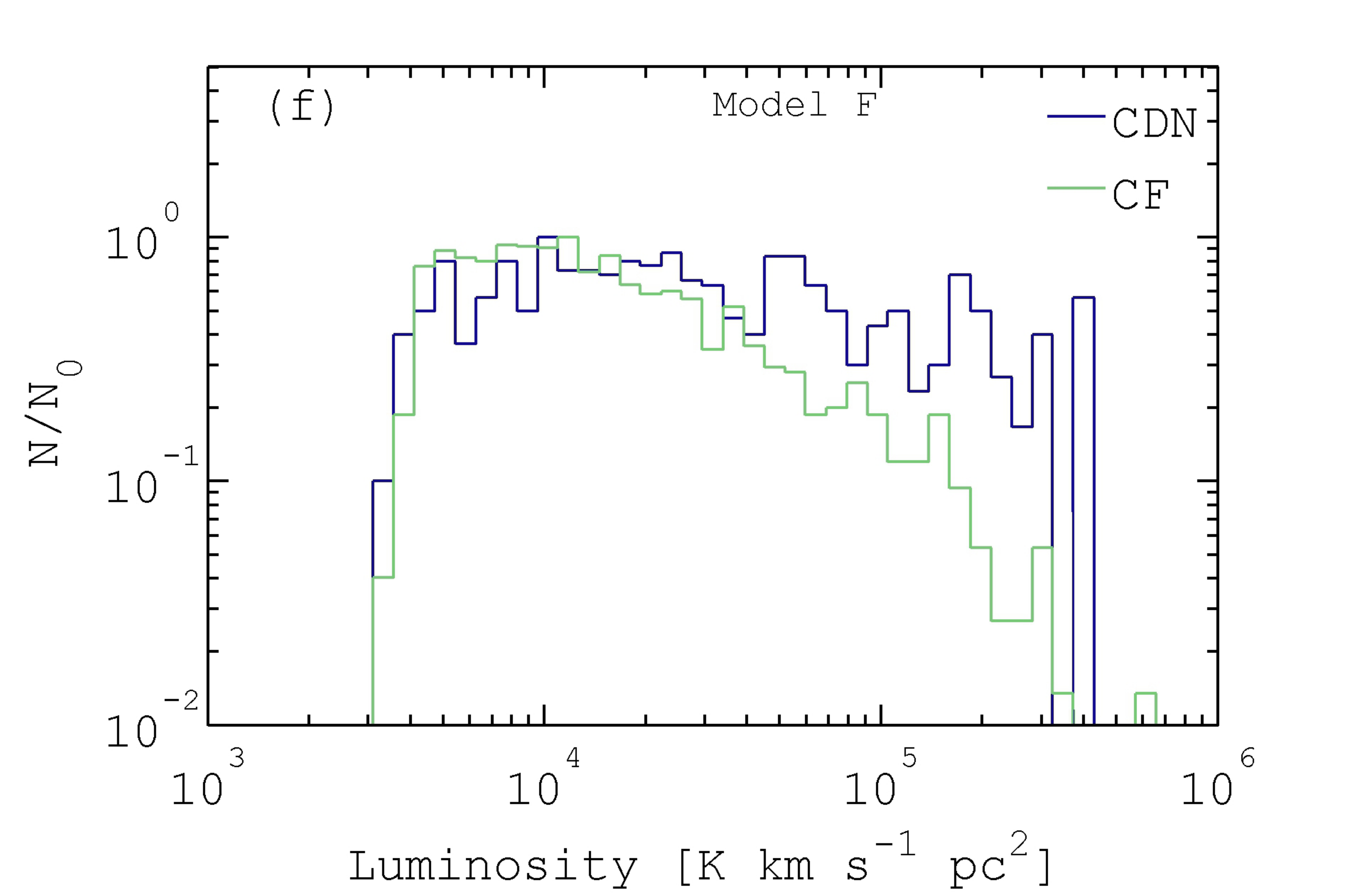}
\caption{
The physical parameters of GMCs obtained using CDN criterion (blue) and CLUMPFIND algorithm (green) for the model F
galaxy (middle panel of Fig.~\ref{fig::galaxies}): size ~(a), mass~(b), virial parameter~(d), velocity dispersion~(e)
and luminosity~(f). The dependence of virial parameter on mass of a cloud is shown in panel~(c). The red dotted and black 
solid lines depicted in panels (c) and (d) correspond to $\alpha=1$  and $\alpha=2$, respectively.
}\label{fig::clouds_par}
\end{figure*}

Number of clouds extracted by using both criteria for the cloud definition described above depends on a choice of threshould. For the fiducial values of threshold $N^{\rm th}_{\rm tot} = 1.9\times 10^{21}$~cm$^{-2}$ and  $T^{\rm th}_{\rm b} = 3$~K we extract $\sim 1000$ isolated clouds in our simulated galaxies. These clouds have the following physical parameters: masses are $\approx 10^4 - 10^7$~\Msun, sizes vary within the range $3 - 100$~pc, one dimensional velocity dispersion is ranged $0.1-10$~km~s$^{-1}$, mean surface densities are $\sim 60-300$~\Msun~pc$^{-2}$, and luminosities in CO lines are $10^3-10^7$~\Lum. These parameters depend slightly on the galactic morphology. Figure~\ref{fig::clouds_par} shows the distributions of these physical parameters for the cloud population in the Milky Way type galaxy~(Model F) for both techniques of the cloud definition.  

For{ the} CF method { the} physical paramters of clouds, i.e. masses, 1D line-of-sight velocity dispersion (LOSVD), total luminosity and cloud sizes, are calculated using the prescriptions from \citet{1994ApJ...428..693W}.

For the CDN approach one dimensional velocity dispersion of { a} cloud is calculated according to 
\begin{equation}\label{eq::veldisp}
\Oo  \sigma_v = \frac{1}{\sqrt{3}} \sqrt{ \sum({\bf u} - {\bf u}_{\rm c})^2 }\,,
\end{equation}
where $\bf u_{\rm c}$ is cloud center mass velocity vector, $\bf u$ is cloud velocity vector. Such approach is widely used to extract clouds in numerical simulations~\citep[see e.g.][{ and references therein}]{2013ApJ...776...23B,2014MNRAS.439..936F}. A cloud size is calculated according to \mbox{$\Oo R = \sqrt{\frac{A} {\pi}}$}, where $A$ is  cloud surface in pc$^2$.

The ratio between the kinetic energy and gravitational energy is commonly used to specify deviation from the virial state  of a cloud under assumption of the constant density distribution~\citep{1992ApJ...395..140B}:
\begin{equation}\label{eq::vir_par}
 \Oo \alpha = \frac{5\sigma^2_{\rm v} R_{\rm cl}}{G M_{\rm lum}} \approx \frac{1161 \sigma^2_{\rm v} R_{\rm cl}}{M_{\rm lum}}\,,
\end{equation}
where $\alpha$ is cloud virial parameter, $R_{\rm cl}$ is cloud size in parsecs, $M_{\rm lum}$ is mass of cloud in Solar units adopted from its CO luminosity using $X_{\rm CO}$ conversion factor
\citep{1978ApJS...37..407D}:
\begin{equation}\label{eq::mcl_lum}
\Oo M_{\rm lum} = \frac{4.4L_{\rm CO}X_{\rm CO}}{2\times10^{20}}\,.
\end{equation}
It is easy to see that the luminosity mass $M_{\rm lum}$ and the virial mass $M_{\rm vir}$ of a cloud are generally not equal to each other:
\begin{equation}\label{eq::mvir_mlum}
M_{\rm vir} \approx \alpha M_{\rm lum}\,.
\end{equation}
Middle panels in Fig.~\ref{fig::clouds_par} present the parameter $\alpha$ for cloud population in the  Milky Way type galaxy. It is clearly seen that the majority of molecular clouds tend to be in the virial equilibrium. Note that  the distribution of such quasi-virialized objects is close to{ the} uniform distribution as a function of  cloud mass. The physical parameters obtained for our models of galaxies are in agreement with other recent numerical simulations \citep{2009ApJ...700..358T,2011ApJ...730...11T,2013MNRAS.428.2311K}. Such result is likely to be a reflection of the turbulent energy distribution in the entire galactic disk \citep[detailed study see in][]{2014ApJ...784..112K}.

{ The} distributions of cloud masses { obtained by using the CF and CDN methods are} slightly different: { in the former} we extract smaller and less massive clouds than the ones extracted in the { latter}~\ref{fig::clouds_par}a,b). Moreover, the mass range for the CF sample of clouds is rather wide. The reason of this  is clearly seen in the Fig.~\ref{fig::clouds_clouds}: large structure extracted by{ the} CDN method are divided into several smaller clouds { when} the CF technique{ is used}~(see Fig.~\ref{fig::clouds_clouds}). We discuss several dynamical and methodological effects related to this issue in the further paragraphs.

The 1D velocity dispersion of clouds defined by Eq.~\ref{eq::veldisp} is unlikely to provide a good description for observed line-of-sight velocity dispersion, making it higher at least for extragalactic GMCs. Moreover, the regular quasi-circular motion of giant clouds around galactic center leads to overestimation of velocity field within a cloud about 1-2~\kmps. But this effect is significantly smaller than the 1D LOSVD value. Spectral resolution 0.5~\kmps allows us to distinguish  internal structure of large clouds and measure the 1D LOSVD { within} 1-2~\kmps accuracy~(see Fig.~\ref{fig::clouds_par}e). Using{ the} CF method we extract more homogeneous cloud sample which { has} smoother (without many local peaks) distribution. In any case both methods provide more or less similar shape of the velocity dispersion distribution functions, which are close to the observable ones~\citep[see e.g.][]{2010ApJ...723..492R}. It seems that the CF method splits large clouds into smaller ones due to the { complex} velocity { structure}, which mainly takes place in colliding and tangent gaseous flows or, in general, turbulent regions. Note a remarkable difference between clouds extracted by the CF and CDN methods: large clouds found by the CF method are isolate lumps located in calm environment, whereas large clouds extracted by using the CDN criterion mostly represent dynamically interacting structures.

\section{Scaling relations analysis}\label{seq::scaling_relations}

In this section we discuss the scaling relations for GMCs extracted according to two criteria of the cloud definition for the galaxy models{ described above}. { In the Table~\ref{tab::tabular3} we collect all indices and normalizations for{ the} scaling relations{ obtained in the analysis of} { the models of} galaxies { considered here}. It seems that there is no strong variation of the GMCs scaling relations { obtained by using the CDN method for} galaxies with different morphology. { This} is considered in Sects.~\ref{sec::ll1}, \ref{sec::ll2}, \ref{sec::ll3} { in detail. A role} of the  environment { in the galactic disk} on the GMCs parameters is discussed in the Sect.~\ref{sec::mass_spectra}.}

The statistical relations for the three models of galaxies considered { here} are presented in Figs.~\ref{fig::ll1}, \ref{fig::ll2}, \ref{fig::ll3} and described in the corresponding subsection. The top row of the panels in each figure shows{ the} relations obtained{ by} using the CF method, and the bottom one shows{ the} relations based on the CDN criterion.

\begin{table*}
\caption[]{
The parameters of the scaling relations for GMCs in the simulated galaxies.}
\begin{center}
\begin{tabular}{lcccccc|ccccccccc}
\hline
Model (Morphology) - CD & \multicolumn{2}{c}{$\sigma_{\rm v} = A_1 R_{\rm cl}^{\beta_1}$} & \multicolumn{2}{c}{$M_{\rm vir} = A_2 L_{\rm cl}^{\beta_2}$} & \multicolumn{2}{c}{$L_{\rm cl} = A_3 R_{\rm cl}^{\beta_3}$} \\
          & $A_1$   & $\beta_1$  & $A_2$ & $\beta_2$  & $A_3$ & $\beta_3$ \\
          & \kmps pc$^{-\beta_1}$   &    & \Msun (\Lum)$^{-\beta_2}$ &   & \Lum pc$^{-\beta_3}$ &  \\
\hline
B (No structure) - CDN &  $0.94\pm0.25$  &  $0.66\pm 0.17$ &  $795 \pm 151$  & $0.64 \pm 0.31$  & $200 \pm 45$ 	&  $1.72 \pm 0.13$  \\
B (No structure) - CF &  $0.97\pm0.12$  &  $0.49\pm 0.1$ &  $15.8 \pm 2.2$   & $1.56\pm 0.11$   & $1000 \pm 38$ 	&  $1.02 \pm 0.2$  \\
F (Milky Way like) - CDN &  $0.35\pm0.21$  &  $1.11\pm 0.18$ &  $16 \pm 8.1 $   & $1.04 \pm 0.28$   & $202 \pm 29$ 	&  $1.73 \pm 0.11$  \\
F (Milky Way like) - CF &  $0.57\pm 0.18$  &  $0.69\pm 0.12$ &  $13.1 \pm 3.6$   & $1.47 \pm 0.21$   & $630 \pm 44$ 	&  $1.28 \pm 0.19$  \\
H (Flocculent) - CDN &  $0.87\pm0.24$  &  $0.76\pm 0.16$ &  $1584 \pm 212$   & $0.62 \pm 0.29$   & $156 \pm 51$ 	&  $1.68 \pm 0.3$  \\
H (Flocculent) - CF &  $0.84\pm 0.9$  &  $0.54\pm 0.09$ &  $16.2 \pm 3.1$   & $1.56 \pm 0.1$   & $1000 \pm 35$	&  $1 \pm 0.21$  \\
\hline
\end{tabular}\label{tab::tabular3}
\end{center}
\end{table*}

\subsection{Velocity dispersion - size relation}\label{sec::ll1}

\begin{figure*}
\includegraphics[width=1.0\hsize]{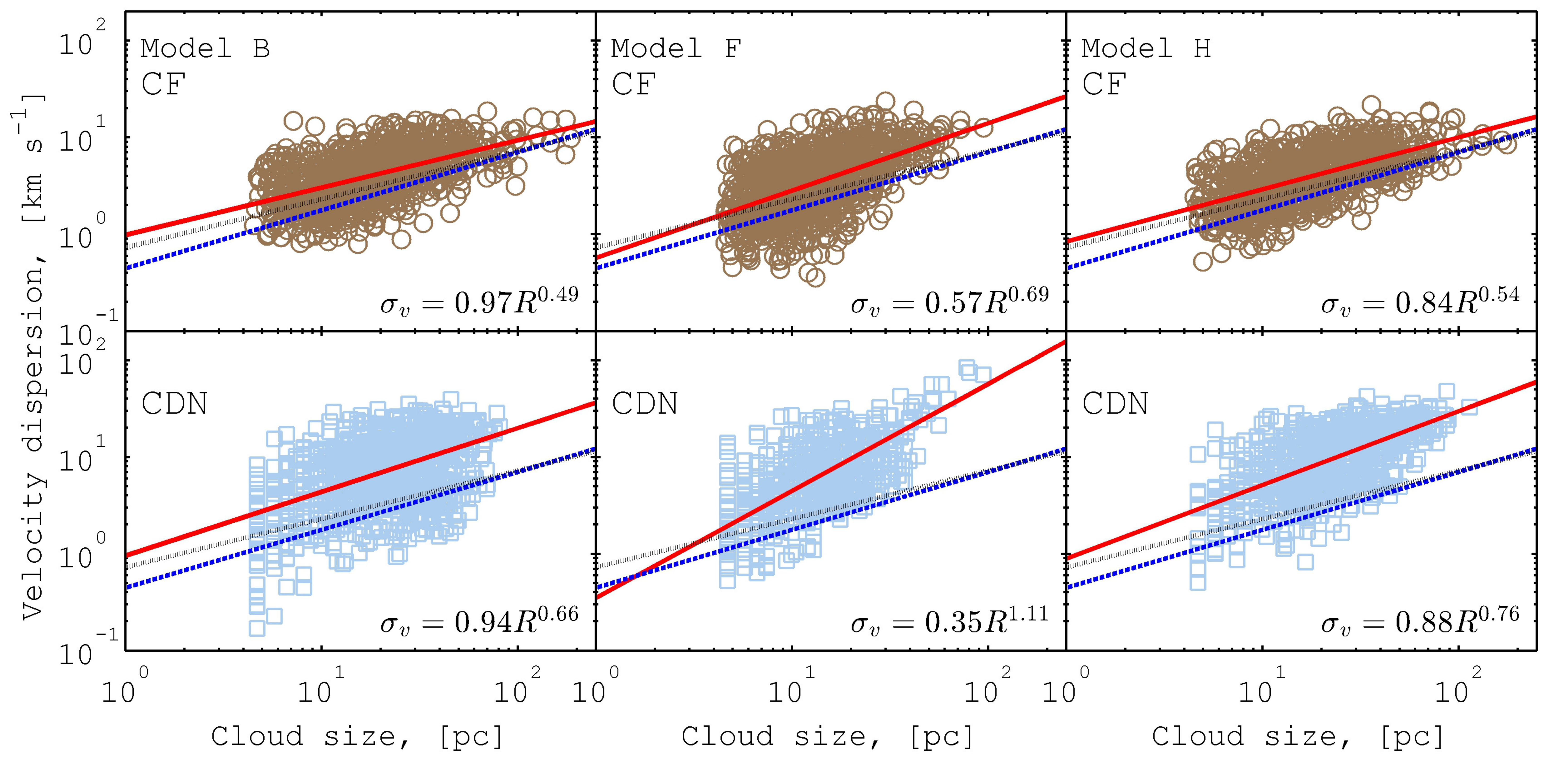}
\caption{
The 'velocity dispersion - cloud size' relation obtained for the CF algorithm (top row of panels) and the CDN criterion (bottom row 
of panels). Left column of panels corresponds to the model of galaxy without spiral structure (model B), middle column presents the
relation for the MW-like galaxy model (model F) and the right column shows the relation for galaxy with flocculent structure (model 
H, see Fig.~\ref{fig::galaxies}). The solid red line is a power-law fit for the simulated data (the corresponding formula is shown in 
the right bottom corner). The dashed blue line corresponds to the fit for the data taken from~\citet{2008ApJ...686..948B}.
The dotted black line shows the fit to the data on the Milky Way clouds got by~\citet{1987ApJ...319..730S}.
}\label{fig::ll1}
\end{figure*}

\begin{figure*}
\includegraphics[width=1.0\hsize]{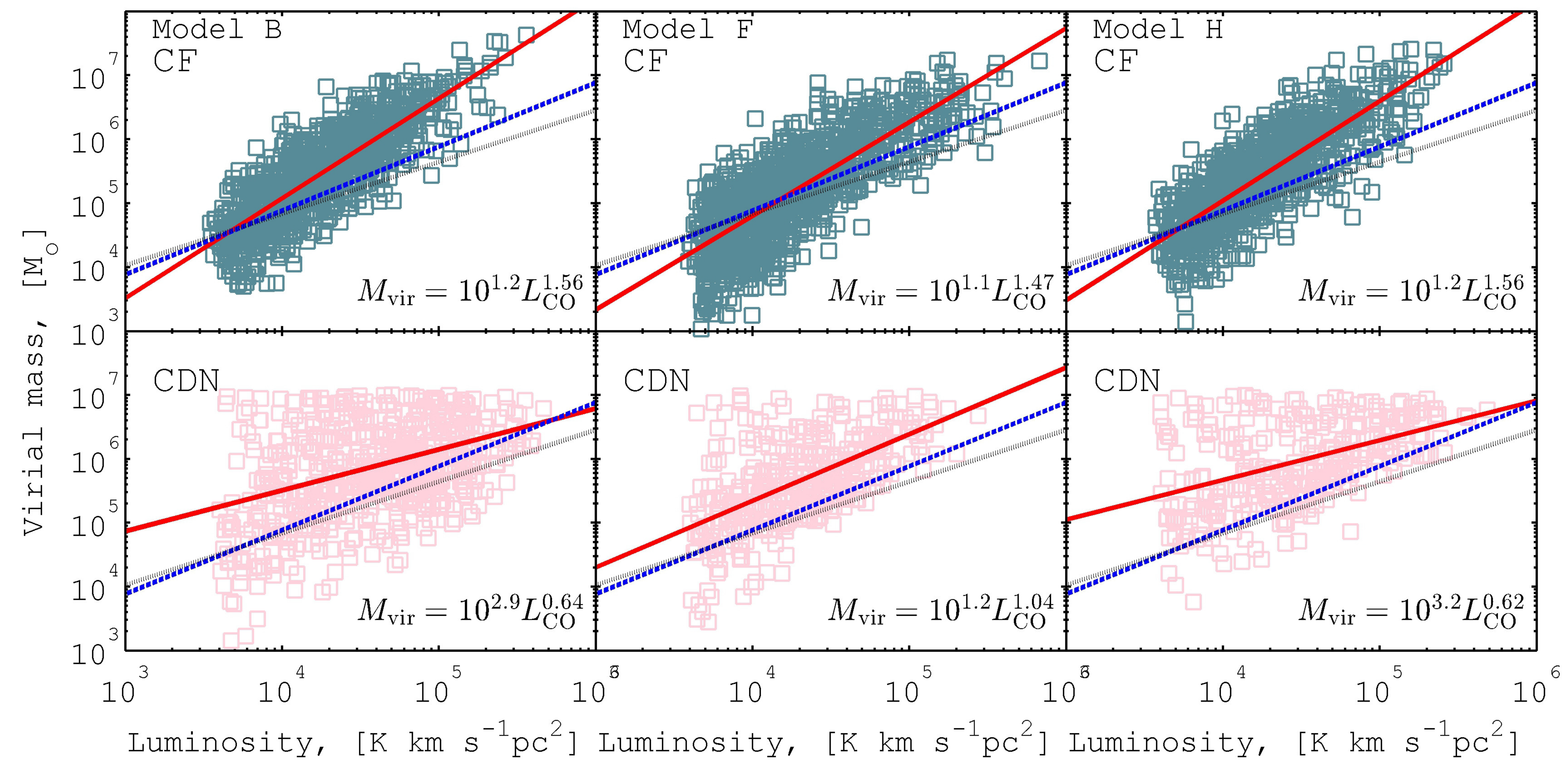}
\caption{
The 'cloud virial mass - luminosity' relation. The other notations are the same as in the Fig.~\ref{fig::ll1}.
}\label{fig::ll2}
\end{figure*}

\begin{figure*}
\includegraphics[width=1.0\hsize]{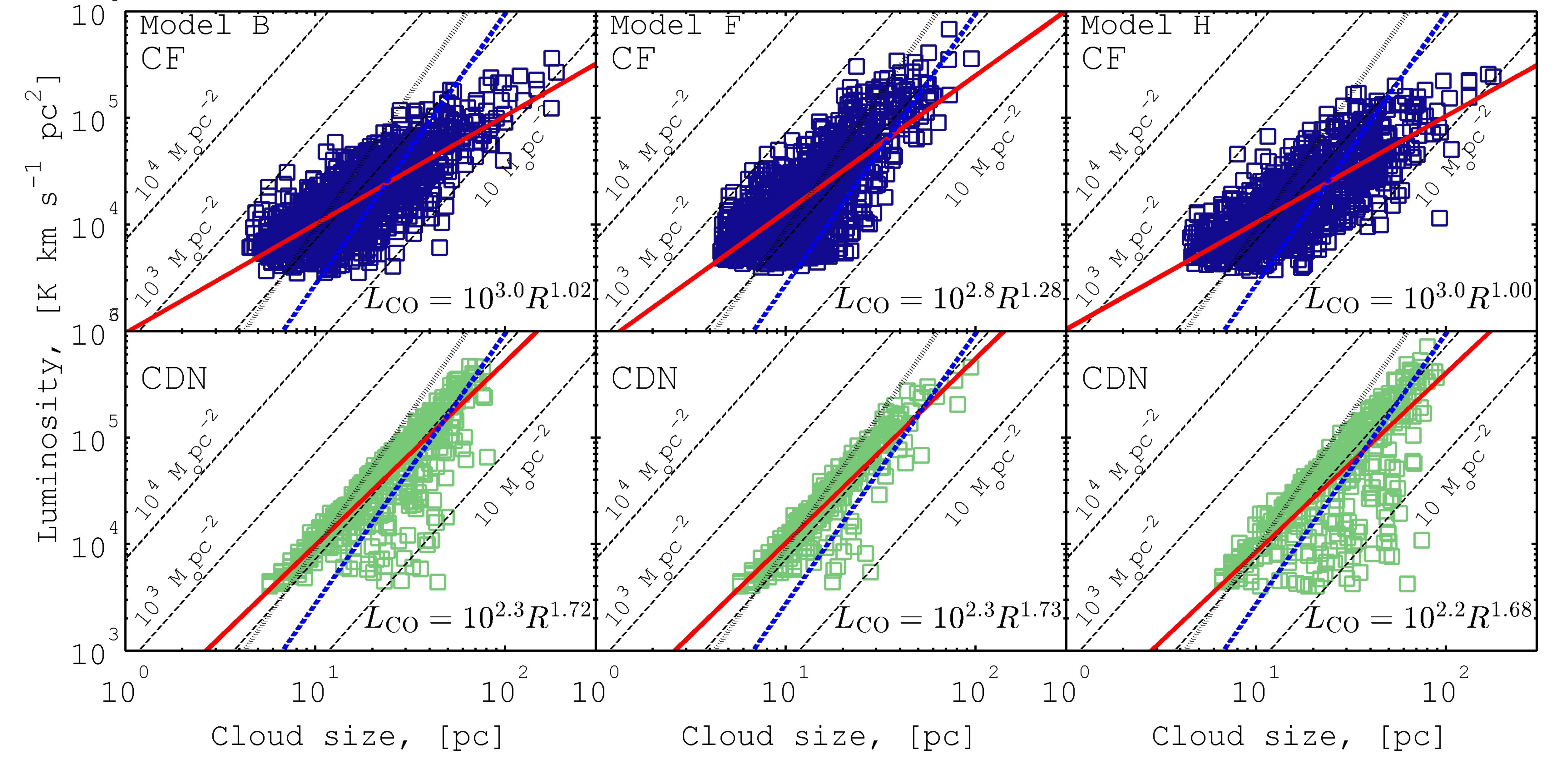}
\caption{
The 'luminosity - cloud size' relation. The dotted lines correspond to the cloud surface density according to Eq.~\ref{eq::mcl_lum} 
for the constant conversion factor $X_{\rm CO} = 2\times10^{20}$~\Xcounit. The other notations are the same as in Fig.~\ref{fig::ll1}. 
}\label{fig::ll3}
\end{figure*}

Fig.~\ref{fig::ll1} { shows} that the clouds extracted according to the CDN criterion have higher velocity dispersion with the mean value $\sim 8-20$~\kmps{ compared to that} { obtained} for the CF { one, in this case} the mean { value of velocity dispersion} decreases to $\sim 1-5$~\kmps. { One can note that} the observational fits{ for the Milky Way galaxy and others}~\citep{1987ApJ...319..730S,2008ApJ...686..948B} { are in} better agreement { with the} relations obtained for the CF sample of clouds. The relations for the clouds extracted by the CDN method show { significant} deviation from the observational fits.

Clouds have extremely complicated shape { and consist of crossed and}  elongated structures (see Fig.~\ref{fig::clouds_clouds}). So that high total hydrogen column density at the periphery of clouds can be a geometrical effect when the line-of-sight goes along the { largest} dimension of the cloud. So the use of the CDN criterion can result in incorrect estimation of cloud sizes and overestimation of their column density and velocity dispersion (Fig.~\ref{fig::clouds_par}).

{ From the bottom panels} in Fig.~\ref{fig::ll1}{ one can conclude} that for the adopted{ here} column density threshold { the clouds extracted by using the CDN criterion} { hold} gas with higher velocity located at their periphery. Such intercloud (diffuse) gas can contain significant molecular fraction.{ Note that i}n some recent observations extended structures with significant molecular fraction are found around molecular clouds \citep{2015AJ....149...76C}. The velocity dispersion of these structures is higher than that in the clouds. This can be considered as an evidence that molecular gas can exist in two phases: clumpy phase, which is organized in molecular clouds, and diffuse one, which is located in extended structures around clouds.

CO molecules are efficiently formed only in the dense shielded environment and are destroyed due to heating { and photodissociation} by the stellar feedback. Using the CDN criterion total hydrogen column density { has a} deposit from not only central dense molecular regions, but also peripheral  parts of a cloud{ that can mainly contain atomic hydrogen} and{ even} some intercloud starforming regions, where young stars have already existed. If we extract clumps brightly emitting in CO lines, then the low density HI gas at the periphery is excluded from the consideration. 

One can see that the indices in the power-law relation $\sigma_{\rm v} - R_{\rm cl}$ for the models of galaxies with more pronounced structures significantly differ from that for the observational fits ~\citep{1981MNRAS.194..809L,1987ApJ...319..730S,2008ApJ...686..948B}. This deviation takes place for both threshold criteria, but it is smaller for{ the} CF method. This can be explained by the fact that { using the} CDN criterion we can consider gaseous structures which are not really associated with  clouds. So that{ gaseous} flows at the outskirts of cloud are added to the internal turbulence motions { of this} cloud, hence, the numerical ratio between velocity dispersion and size of cloud becomes higher. We mentioned above that the CF approach is a sharper 'filter' for molecular clouds than the CDN one, and our cloud samples based on the CO datacubes data demonstrate{ the} statistical relations closer to the observed ones. Thus, in our simulations the first Larson's scaling relation is better reproduced for PPV data.

\subsection{Virial mass - luminosity relation}\label{sec::ll2} 

The 'virial mass - luminosity relation' reflects a suggestion that GMCs state is close to the virial equilibrium. Figure~\ref{fig::ll2} shows the correlation between virial mass $ M_{\rm vir}$~(see Eq~\ref{eq::mvir_mlum} ) and total luminosity of{ the} extracted molecular clouds for three models of galaxies. One can see { significant} scatter of{ the} physical parameters for the simulated clouds around the observational fits. { Similar to} the 'velocity dispersion - size' relation{ one can see here also that} the scatter for the CDN criterion is larger than that for the CF one, especially this is remarkable for low surface luminosity clouds ($L_{\rm CO}\simlt 10^4$~\Lum). In general, for the same luminosity value the virial mass of clouds obtained for the CDN criterion is systematically greater than that for the CF one. That can be explained by that high virial masses have large clouds formed in collisions of smaller ones, such massive clouds are mainly associated with spiral arms and/or bar. During collision of clouds, molecules can be destroyed, but the shock waves cannot ionize gas (or such gas recombines rapidly). So that significant part (on mass) of such clouds is locked in the warm atomic hydrogen phase. Then, using{ the} CDN criterion we obtain clouds with high total hydrogen column density, where the deposit of atomic hydrogen to the column density is dominant or very significant.

The CO brightest clouds are really molecular ones and they are probably to belong to older cloud population: a gas inside them has to become molecular~\citep{2010MNRAS.404....2G}. Whereas the darkest (massive) clouds in CO line are believed to be either young population of massive clouds, in which atomic hydrogen hasn't yet transformed into molecular, or maybe pseudo-virialized structures, which consist of a group of small molecular clouds 'bounded' by atomic intercloud and/or more diffuse molecular medium~ (see~\ref{fig::clouds_clouds}). Such structures can appear in dense environment, e.g. spiral arms, bar, central parts of a galaxy, where a chosen threshold is low enough to extract separate clouds and leads to merger small clouds into larger structure. The use of{ the} CF criterion provides more reasonable cloud sample and doesn't lead to extracting such large gaseous structures. So that the scatter for the sample of clouds obtained for the CF criterion is much smaller that for{ the} CDN one.

For the CDN criterion the slope of the fit for the simulated sample of clouds is flatter than that for the observational data (see Fig.~\ref{fig::ll2}). Obviously, it comes from the excess of massive clouds with low surface luminosity. Using the CF approach the picture for all three models of galaxies shows opposite behaviour. The slope becomes steeper than that obtained in the observations. Above one can see that the CF method usually leads to splitting large structures into smaller ones due to systematic LOS velocity variations for a given structure~(see Fig.~\ref{fig::clouds_clouds}), that reveals in both size and mass distributions~(see Fig.~\ref{fig::clouds_par}). Thus, we have a relatively large subsample of small clouds, which cannot be resolved (in space and/or in the LOS velocity coordinate) in observational data.

One can suppose that if{ a} large gaseous agglomeration in the vicinity of spiral arms or galactic centre { is} splitted into several isolated clouds, then{ the} number of massive and bright clouds becomes lower, whereas smaller clouds are more numerous. Thus, the slope of the fit can become flatter. So that the increase of threshold value is likely to result in better match with observations.

\subsection{Luminosity - size relation}\label{sec::ll3} 

Originally Larson (1981) found the 'mass-size' relation for the Galactic molecular clouds: $M_{\rm cl} \propto R^2_{\rm cl}$. This relation can be interpreted as  molecular clouds have the same (constant) surface density. Here we use another form of this relation, namely 'luminosity - size', because it includes at least one observable value. Note that mass of a cloud can be easily found from luminosity using conversion factor $X_{\rm CO}$ according to Eq. (\ref{eq::mcl_lum}), such re-calculation doesn't affect on the slope of the scaling relation in case of the constant conversion factor.

Figure~\ref{fig::ll3} shows the relation for three models of galaxies. It is clearly seen that for all models the surface density of clouds is locked within interval $\sim 10-1000$~\Msun~pc$^{-2}$. This surface density { range} is rather universal within $M=10^3-10^7$~\Msun for all galaxy models. For the CDN criterion one can note { substantial} scatter of the cloud parameters below the critical value of the surface density $\approx 10^2$~\Msun pc$^{-2}$~(see bottom row in Fig.~\ref{fig::ll3}). This is just a reflection of existence of the dimmer parts of the clouds. It seems that the strong limit on the maximum value of the surface density can be { interpreted as a result of} ongoing star formation, which prevents the formation of more dense clouds. Molecules in such clouds are destroyed immediately due to photodissociation by UV radiation from  newborn stars. However, such picture cannot be supported by the analysis of the clouds extracted by CF method (see top row in Fig.~\ref{fig::ll3}). It is possible that the brightness of large clouds becomes lower than { that} expected { due to} the shielding effects. Note that the optical depth effects become important when the value of the column density exceeds $\sim 2\times 10^{21}(T/10^3)^{-1}$~cm$^{-2}$ \citep[e.g.,][]{1979ApJS...41..555H} and dense parts of clouds become dimmer in the CO lines.

Note that in our simulations gas number density can be high as 2000-3000~cm$^{-3}$. However, even in such dense medium a star does not form with necessity, because a gas can be in the equilibrium with the surrounding medium. Such picture is usually taken place in small clouds. So that sometimes one can find rather small clouds (see Figs.~\ref{fig::clouds_par}a,b) with large amount of molecular gas, and these clouds appear to be brighter than it is expected from third Larson's relation. Thus, in our calculations the clouds surface density is expected to be not always constant that reflects that in our model there is no gas density threshold for the star formation process.

\begin{table}
\caption[]{The power-law indices of the scaling relations found in the Model F for several values of CO spectral line resolution $\delta v$. The observational fits are given in the top frame of Table.}
\begin{center}
\begin{tabular}{lcccccc|ccccccccc}
\hline
Model / Observations &$\delta v$ & $\beta_1$  & $\beta_2$ & $\beta_3$ & \\
\hline
 & \kmps &    &   &   & \\
\hline
\citet{1981MNRAS.194..809L} &  - & 0.38 & - & -\\
\citet{2008ApJ...686..948B} &  & 0.5 & 0.81 & 2.55\\
\citet{1987ApJ...319..730S} &  & 0.6 & 1 & 2.54\\
\citet{2010ApJ...723..492R}  & 1 & - & - & 2.36\\
\hline
Model F & 0.5 & 0.69 & 1.47 & 1.28 \\
Model F & 1    & 0.65 & 1.34 & 1.74\\
Model F & 5    & 0.6   & 1.19 & 2.41\\
\hline
\end{tabular}\label{tab::tabular2}
\end{center}
\end{table}

\subsection{Variation of the spectra resolution}

Spatial resolution in numerical simulation plays a significant role in understanding of the internal properties and basic physical parameters of GMCs. \citet{2014MNRAS.439..936F} reported that the variation of spatial resolution  strongly affects on the properties of cloud populations. At the same time, results of the PPV data cube { analysis} can depend on spectral resolution. In our previous simulations of synthetic spectra { the} velocity resolution { equals} $\delta v =0.5$~\kmps, which is quite high for extragalactic observations. Although this value is comparable to that used in several studies \citep[e.g.,][]{2013MNRAS.436..921T}, most of the recent extragalactic surveys in molecular lines have been done with much lower spectral resolution \citep{2003ApJS..149..343E,2013ApJ...772..107D,2013ApJ...779...42S,2014A&A...565A..97C}.

To check whether spectral resolution affects on the scaling relations, we calculate and analyze the PPV data with lower spectral resolution $\delta v = 1$ and $5$~\kmps for the model F. In Tab~\ref{tab::tabular2} we show the power-law indices for the scaling { relations} with different resolution values, we also combine the indices obtained in several observations with brightness temperature threshold{ equal to} 1~K. We argue that the noticeable variations of the indices with $\delta v$ are due to that for lower spectral resolution small clouds are combined into larger ones on the line of sight when their relative motion and velocity dispersion is lower or comparable with the spectral resolution. Here we only report that there is a dependence of the cloud population characteristics on spectral resolution. An accurate quantitative consideration of this effect requires further detailed study.

\begin{figure}
\includegraphics[width=1\hsize]{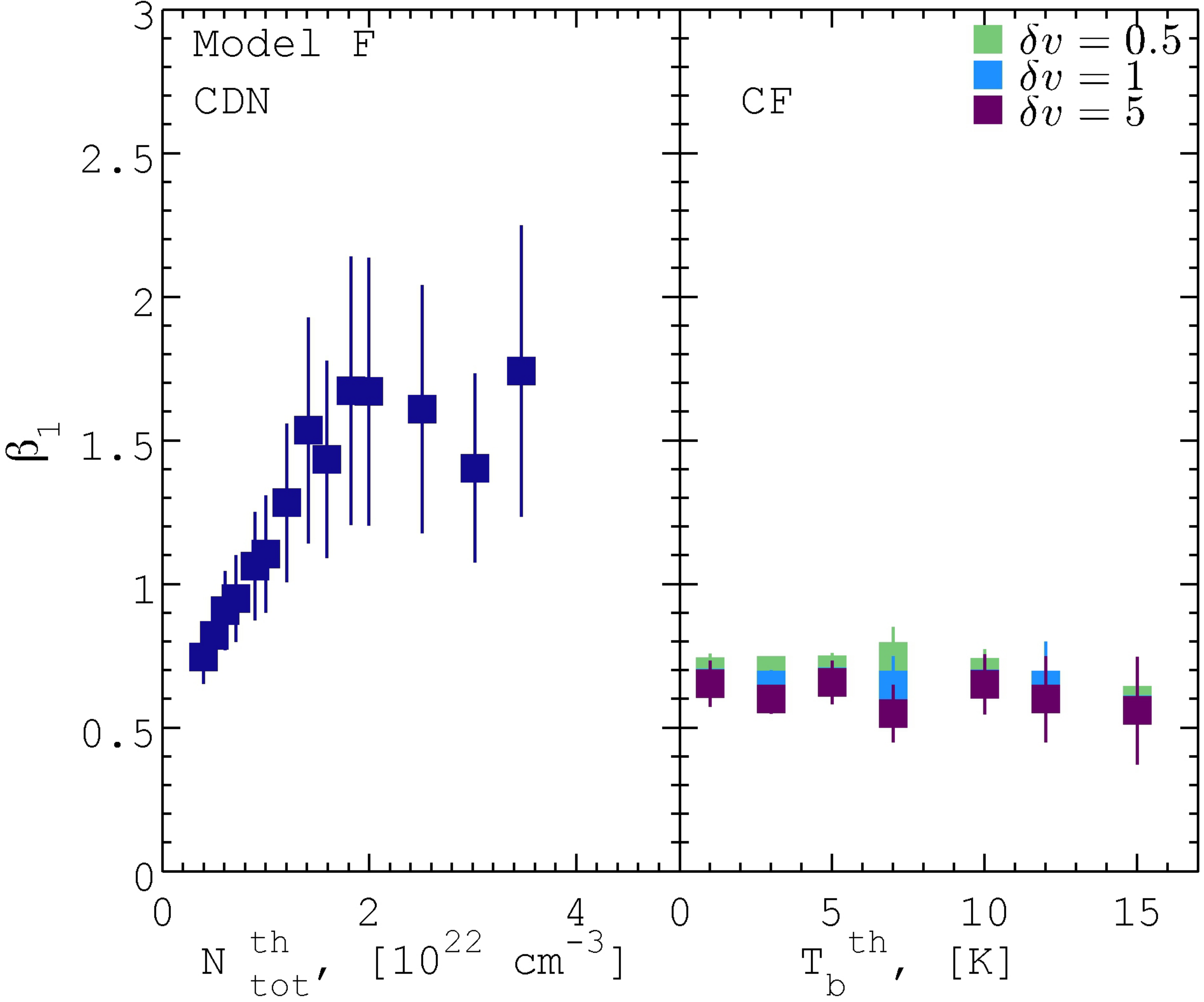}
\includegraphics[width=1\hsize]{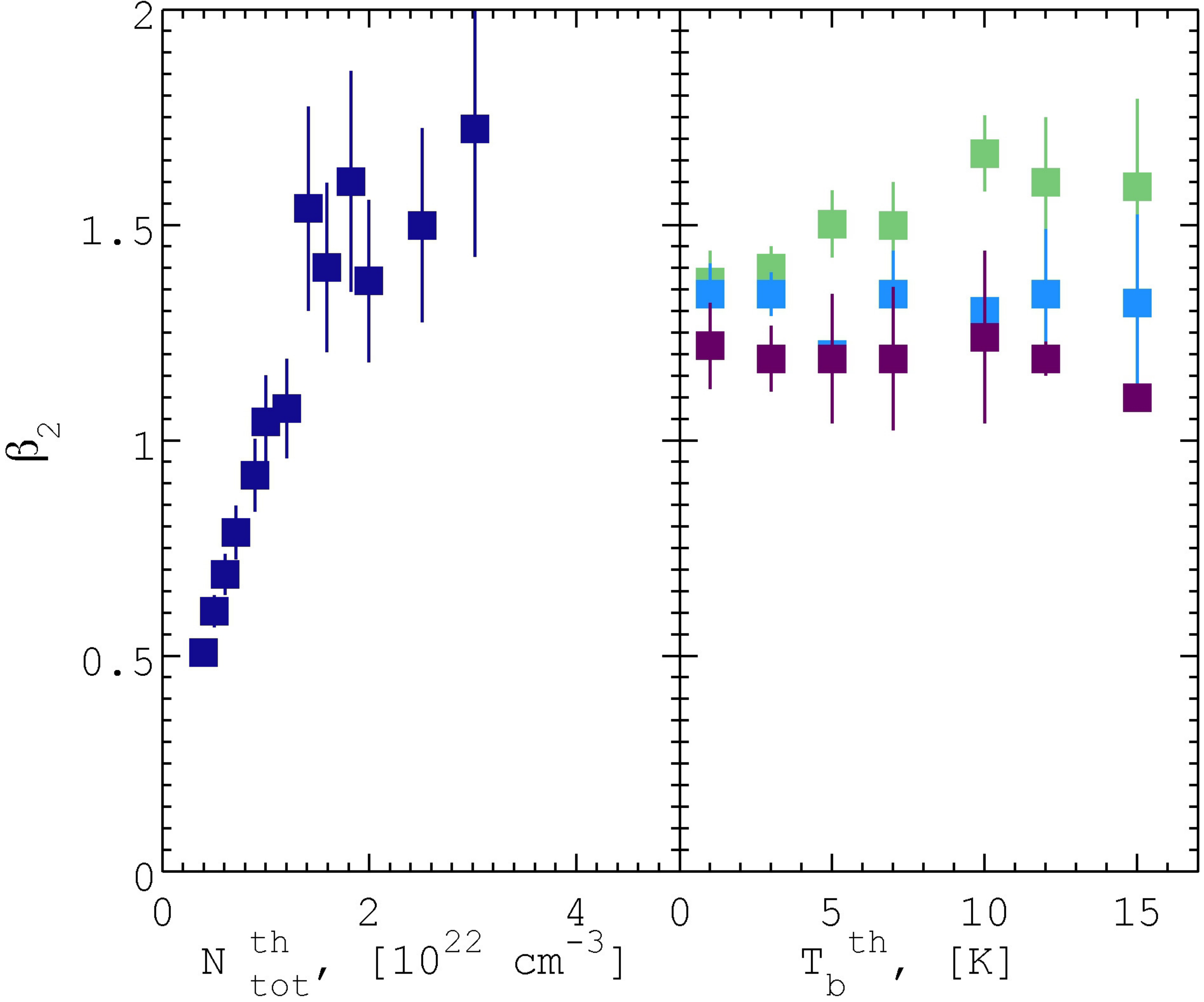}
\includegraphics[width=1\hsize]{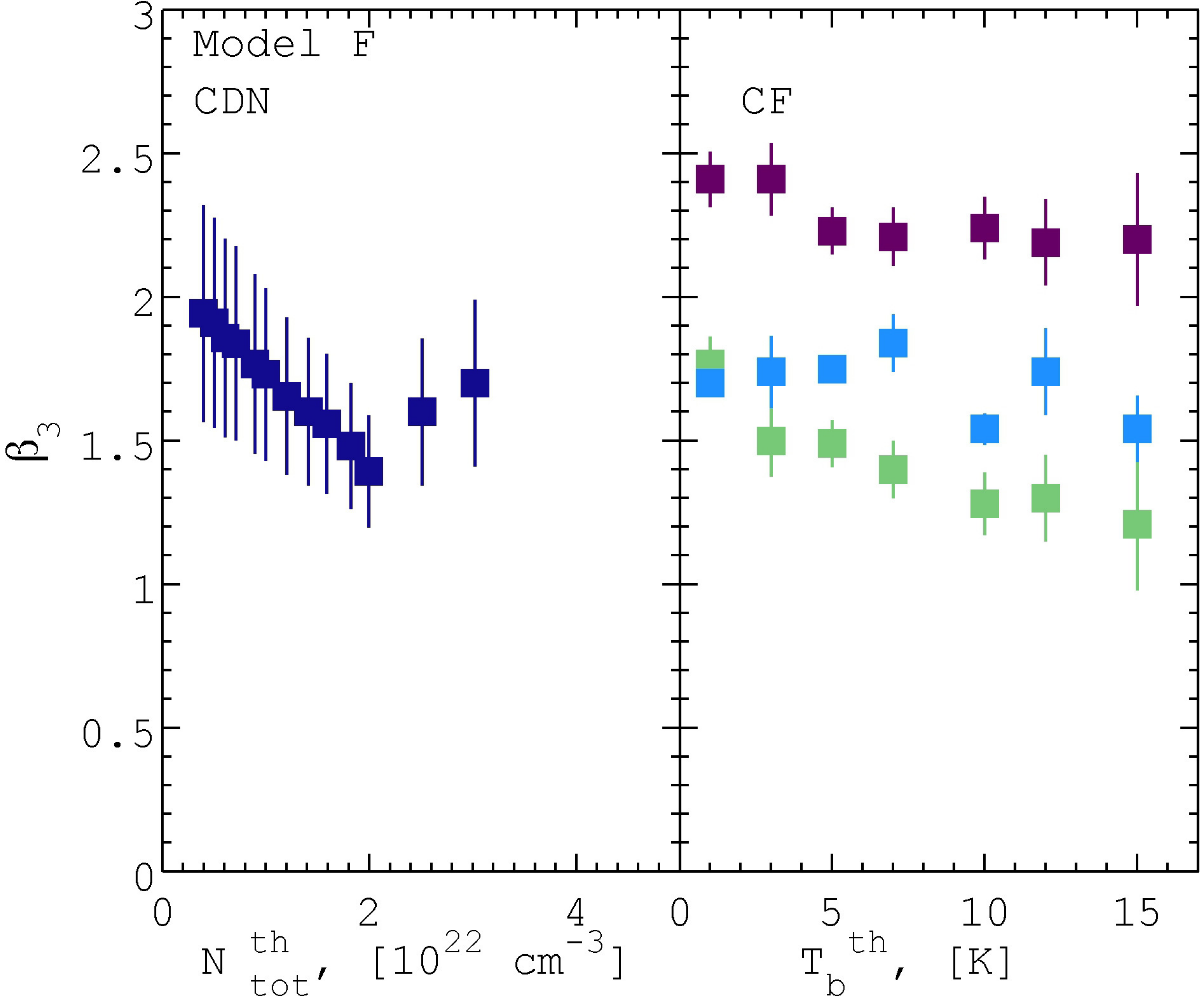}
\caption{The power-law indices for the scaling relations, $\beta_i$, $i=1,2,3$, as function of the total column density threshold $N^{\rm th}_{\rm tot}$ for clouds extracted using CDN approach~(left panels). In the right panels the indices $\beta_i$ are shown as a function of the brightness temperature $T^{\rm th}_{\rm b}$ for the cloud samples found in analysis of CO line spectra using the CF method for various spectral resolution $\delta v$. }\label{fig::ll4}
\end{figure}

\subsection{Variation of the threshold value}

In the previous subsections we{ have} established that the scaling relations for the cloud ensembles obtained in our simulations are quite similar to these found in the observations. It is interesting to study a dependence of the power-law indices of the relations on  value of threshold. 

We consider the relations obtained by both methods described in Section 3. Here we constrain ourselves to analyse the model of galaxy with the prominent structure -- model F. To do this we vary threshold values in the following ranges: $N^{\rm th}_{\rm tot} = (0.5-4)\times 10^{22}$~cm$^{-2}$ for CDN and $T^{\rm th}_{\rm b} = (1-15)$~K for CF. Using the lower limits{ we extract} clouds with mass less than $10^8$~\Msun, while for the upper values at least 100 molecular clouds remain in the catalogue.

Figure~\ref{fig::ll4} presents the dependence of the power-law indices for three scaling relations on  total column density threshold $N^{\rm th}_{\rm tot}$~(left panels) and brightness temperature $T^{\rm th}_{\rm b}$ threshold~(right panels). The error bars correspond to the data dispersion obtained in { the} $\chi^2$-fitting procedure of the power-law indices for a given threshold. Note that the number of clouds definitely depends on threshold value, but{ it} remains above 100  clouds in order to provide enough objects for statistics.

For low $N^{\rm th}_{\rm tot}$ we extract both extremely large and small clouds. Large clouds consist of a group of small clouds enclosed by the extended diffuse structure, which can be called as the intercloud medium. Such a structure includes both molecular and atomic gas. The increase of the threshold excludes the intercloud medium. So that { for higher} threshold values we extract bright cores of the virialized clouds.  One can see in Fig.~\ref{fig::ll4}~(left panels) that the indices $\beta_i$ for the simulated clouds change dramatically in the case of using{ the} CDN method. Better agreement with{ the} observational data can be found only with relatively low threshold values $\approx 0.5-1\times10^{22}$~cm$^{-2}$. 

Whereas the use of the CF method does not provide any significant variations of the scaling relation indices with threshold (see right panels in Fig.~\ref{fig::ll4}). In this case the cloud populations are vanished from capture of the intercloud medium, because { this} extraction method directly relates to regions with high molecular fraction. This explains { that the} result { remains rather robust} { relative} to variation of  brightness temperature threshold value. The indices obtained by using the CF criterion are close to the observed ones in wide range of threshold{ values}. 
Moreover, one cannot see significant dependence on velocity resolution for the first scaling relation. For the other relations the decrease of resolution leads to{ the} systematic shift of the index values, so that the dependence on threshold value remains more or less flat.

\subsection{Clouds mass spectra}\label{sec::mass_spectra}

The { indices} of the GMCs scaling relations slightly vary for galaxies with different morphology, although both methods of the cloud definition are suffered from so-called environmental effects~(see Figs.~\ref{fig::ll1},~\ref{fig::ll2},~\ref{fig::ll3}). In our case such effects come from remarkable large scale structures like spiral arms and galactic bar.

To check the impact of the galactic environment on the clouds properties we calculate the cumulative mass functions for three types of galaxies, i.e. number of clouds $N$ with masses $M_{\rm cl}$ greater than a reference mass $M'_{\rm cl}$:
\begin{equation}
N(M'_{\rm cl}) = N(M_{\rm cl}<M'_{\rm cl})\,.\label{eq::cnum1}
\end{equation}

The mass spectrum of molecular clouds is usually expressed as a power law{ function}~\citep[see e.g.][]{2007ApJ...654..240R,2012A&A...542A.108G}, however more accurate approach is based on the truncated power law{ shape}~\citep{1997ApJ...476..166W,2014ApJ...784....3C}{ which can be written} in the form:
\begin{equation}
N(M_{\rm cl}<M'_{\rm cl}) = N_0 \left[ \left(\frac{M_{\rm cl}}{M'_{\rm cl}}\right)^{\gamma+1} - 1 \right]\,\label{eq::cnum2}
\end{equation}
where the index $\gamma$ shows how the mass is distributed among the cloud population. 

\begin{figure}
\includegraphics[width=1\hsize]{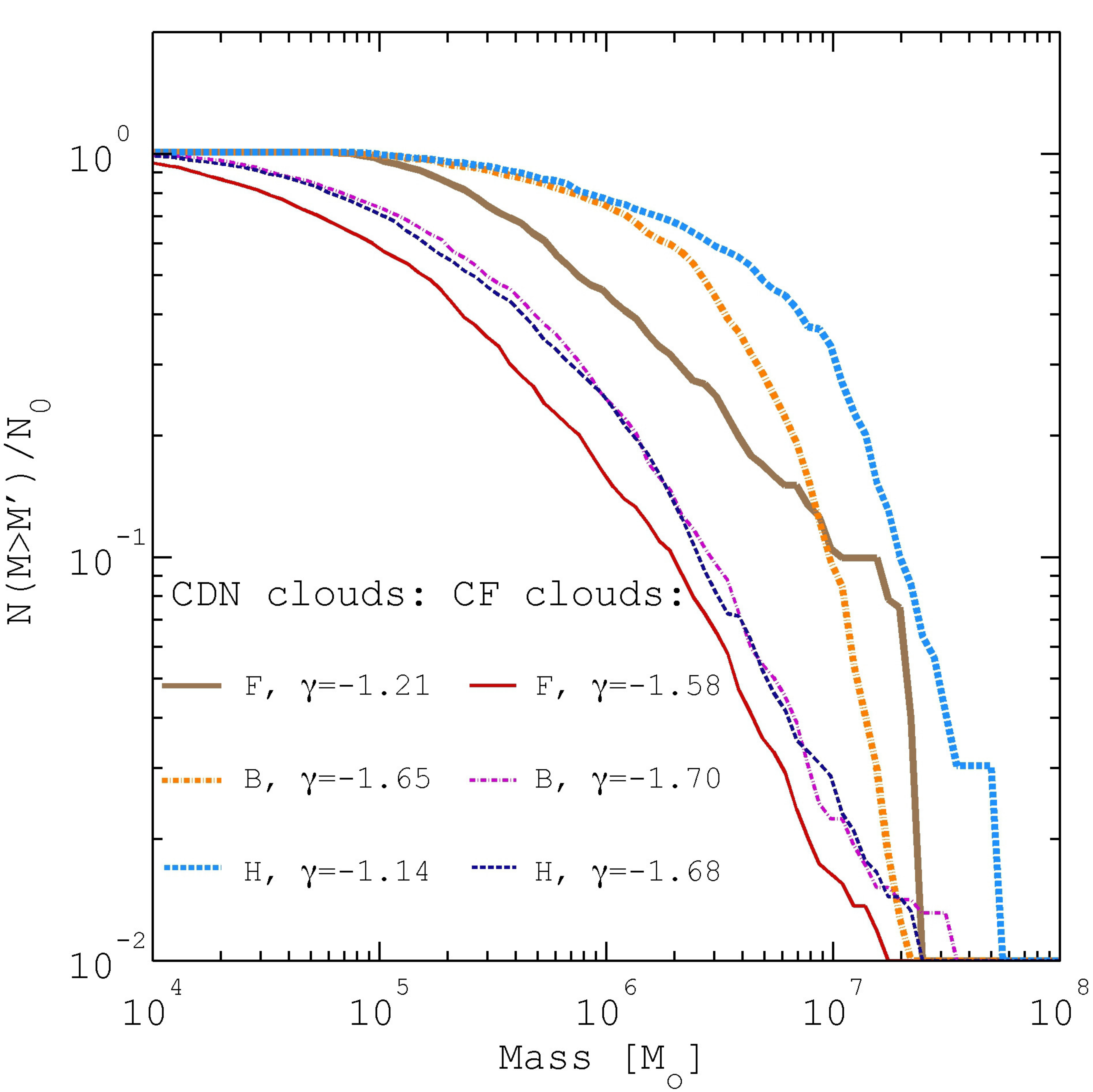}
\caption{
The mass spectrum of GMCs extracted by using{ the} CDN (three top lines) and{ the} CF (three bottom lines) methods in the models of galaxies (Tab.~\ref{tab::tabular1}). The index $\gamma$ for the mass spectrum in the form (\ref{eq::cnum2}) is shown{ in the legend} .
}
\label{fig::mass_spec}
\end{figure}

We compute the fits of the cumulative mass distribution in the form (\ref{eq::cnum2}) (Fig.~\ref{fig::mass_spec}). For all considered models the slope of the mass distribution function $\gamma$~{ (Eq.~\ref{eq::cnum2})} is greater { than} $-2$ that means that large massive clouds dominate in the total GMCs mass budget. One can see that the clouds in the CDN samples have rather steeper mass distribution than that for the CF sample. This demonstrates that most molecular mass tends to be concentrated in less massive clouds in the CF sample than that in the CDN one. { In other words,} small clouds { are} more numerous in the CF sample. That is clearly seen from Fig.~\ref{fig::clouds_par} and even from Fig.~\ref{fig::mass_spec} if one mention that the total masses of extracted clouds obtained using both methods are very close to each other. Such conclusion is general for all three galaxy models~(see Fig.~\ref{fig::mass_spec}). 

{ A remarkable  truncation of the mass distribution is seen for{ the CDN cloud samples} in all models. This can be explained by the engaging of numerous structures above the column density threshold in the dense environment. This suggestion is confirmed by Fig.~\ref{fig::mass_spec} where{ one can see that this} effect is { more clear in}{ the} { galaxy} with the prominent spiral pattern and bar (model F) and { weaker} { in the galaxy} without structure (model B).{ Such s}ignificant truncation does not reflects the physical state of isolated molecular{ clouds}, because{ the} truncation is not detected for{ the CF sample of} GMCs.}

{As { it} was mentioned above~(see Sections~\ref{sec::ll1},~\ref{sec::ll2},~\ref{sec::ll3}), there are not strong variations of { the} scaling relation{ indices on galactic} morphology for { the same} CD criterion. However, the impact of the galactic environment on the GMCs properties{ reveals distinctly}. It is seen in the mass distribution profiles in Fig.~\ref{fig::mass_spec}.} The shapes of the distributions for the CF samples are quite similar to each other: the $\gamma$ { values are in the range $[-1.70;  -1.13]$}, that { once again shows} a homogeneity of these cloud samples. Especially this is remarkable for the galaxies without large-scale pattern: models B and H. Note that for the CDN samples the distributions coincide for $M\simlt 10^6~\Msun$. The mass spectrum in { the} model F systematically differs from the others. { We suggest that { stronger} stellar feedback and compression of GMCs in spiral arms { taken place} in the model F{ substantially} affects  on the GMCs mass distribution. However, { the conformity of the} scaling relations~(see Table~\ref{tab::tabular3}) denotes that GMCs save their internal structure or{ the} CD methods { work in resemling way and as a result} extracted structures~(clouds) { have} rather { close} physical parameters.} Similar influence of large scale structures can be noticed both in numerical simulations of M83 \citep{2014MNRAS.439..936F} and in observations of M51 \citep{2014ApJ...784....3C}. More detailed discussion{ of} such influence on statistical properties of clouds from{ the} observational point of view can be found in \citep{2013ApJ...779...46H}.

\section{Conclusions}\label{sec:summary}

In this paper we have presented a set of the galactic scale simulations of the Milky Way size galaxies of different morphological type: a galaxy without prominent structure, a spiral barred galaxy and a galaxy with flocculent structure. In our models we have taken into account star formation, stellar feedback, UV radiation transfer and non-equilibrium chemical kinetics for CO and H$_2$ molecules. Here we have focused on the statistical properties of molecular clouds obtained by two different extraction methods of { gaseous structures}. The first uses the total hydrogen column density threshold as a marker of the cloud border. The other cloud definition method is based on extraction from { position-position-velocity (PPV)} data cubes for $^{12}$CO~(1-0) line. Using both methods we have studied the empirical scaling relations known as Larson's laws: 'velocity dispersion - cloud size', 'luminosity - cloud size' and 'virial mass - luminosity' relations. Using our simulations we have created the position-position-velocity data cubes for several values of velocity resolution and have investigated how the physical parameters of clouds and the indices of the scaling relations depend on spectral resolution. Our results can be summarized as follows:

\begin{itemize}
 \item the number of spatially resolved molecular clouds in the simulations slightly depends on galactic type and equals 
  $\sim 10^3$;  size, mass, luminosity and other physical properties of giant molecular clouds obtained in { the} simulations are 
  close to { the} observed ones in our and nearby disk galaxies; { note that }the physical parameters of clouds depend on the cloud definition 
  method~(see Figs.~\ref{fig::clouds_clouds},\ref{fig::ll1},\ref{fig::ll2},\ref{fig::ll3}).

 \item the diffuse (intercloud) gas can be catched using total column density as threshold in the extraction of clouds, especially this 
  can be significant in dense large scale structures, e.g. within spiral arms or bar; such diffuse gas has higher velocity dispersion and
  lower{ CO line} brightness in comparison with  other cloud material, so that { using} this method of extraction we can not exclude overestimation 
  of the 1D velocity dispersion{ due to} the projection effects even at high total column density threshold $N^{\rm  th}_{\rm tot}$
  values; 
  
 \item giant molecular clouds found by { using the} CLUMPFIND (CF for shortness) algorithm have smaller sizes, masses and velocity dispersion than { those} extracted by 
  using total column density as threshold. However, the distributions of virial parameter for both extraction methods show similar
  behaviour~(see Fig.~\ref{fig::clouds_par});
 
 \item numerical models of galaxies with different morphology produce a { substantial number} of rather small GMCs ($R_{\rm
  cl}<20$~pc) which are detectable by various methods considered in the paper~(see Fig.~\ref{fig::clouds_par}a). { This} is more 
  clear for position-position-velocity analysis, where large clouds are splited into smaller ones due to complex kinematics of{ gaseous flows}.
  However the analysis{ of the mass distribution functions} shows that{ the} mass of the{ cloudy phase in galaxies simulated here} is mostly concentrated in large massive clouds~(see
  Fig.~\ref{fig::mass_spec}); 
  
  \item { physical parameters of GMCs depend{ weakly} on  galactic structure, namely, mass, size, luminosity and velocity dispersion are{ locked} in the same ranges for models{ of galaxies} without structure, with prominent spiral pattern and with flocculent patten~(see Figs.~\ref{fig::ll1},\ref{fig::ll2},\ref{fig::ll3}); indeed, we do not see statistically sensible variations of { the} scaling relations in{ the models of} galaxies with different morphology for a given CD criterion~(see Table~\ref{tab::tabular3}); however so-called environmental effects can be clearly seen in the distributions of cloud masses: the mass spectra are steeper in galaxy with the prominent structure~(see Fig.~\ref{fig::mass_spec}).}
\end{itemize}

Thus, we conclude that it is impossible to extract equivalent clouds populations by using two different clouds extraction methods{ considered here}: the first is based on total column density as threshold and the second is utilized the PPV data analysis. Obviously, the comparison between observational and simulated properties of GMCs should be based on the same cloud extraction technique. A significant role of the cloud definition method and selection criteria (e.g. spectral resolution, threshold value etc.) can correspond to a fact that the observable scaling relations for external galaxies might not completely reflect real physical parameters of the ISM cold phase.

\section{Acknowledgments}
We kindly thank our referee, Erik Rosolwski, for thoughtful suggestions that highly improved the quality of the paper. We also thank Marco Lombardi for several stimulating discussions and reading of the early versions of the manuscript. The numerical simulations have been performed at the Research Computing Center~(Moscow State University) under the Russian Science Foundation grant~(14-22-00041) and Joint Supercomputer Center~(Russian Academy of Sciences). This work was supported by the RFBR grants (14-02-00604, 15-02-06204, 15-32-21062) and by the President of the RF grant (MK-4536.2015.2). SAK has been supported by a postdoctoral fellowship sponsored by the Italian MIUR. AMS has been supported by the Ministry of Education and Science of the Russian Federation within the framework of the research activities (project no. 3.1781.2014/K) and Ural Federal University competitiveness increase program. EOV is thankful to the Ministry of Education and Science of the Russian Federation (project 213.01-11/2014-5) and RFBR (project 15-02-08293). The thermo-chemical part was developed under support by the Russian Scientific Foundation (grant 14-50-00043).

\bibliography{mycloudbib}

\begin{thebibliography}{}

\bibitem[\protect\citeauthoryear{{Allen}, {Hogg} \& {Engelke}}{{Allen}
  et~al.}{2015}]{2015AJ....149..123A}
{Allen} R.~J.,  {Hogg} D.~E.,    {Engelke} P.~D.,  2015, \aj, 149, 123

\bibitem[\protect\citeauthoryear{{Asplund}, {Grevesse} \& {Sauval}}{{Asplund}
  et~al.}{2005}]{2005ASPC..336...25A}
{Asplund} M.,  {Grevesse} N.,    {Sauval} A.~J.,  2005, in {Barnes} III T.~G.,
  {Bash} F.~N.,  eds, Cosmic Abundances as Records of Stellar Evolution and
  Nucleosynthesis Vol.~336 of Astronomical Society of the Pacific Conference
  Series, {The Solar Chemical Composition}.
p.~25

\bibitem[\protect\citeauthoryear{{Bakes} \& {Tielens}}{{Bakes} \&
  {Tielens}}{1994}]{1994ApJ...427..822B}
{Bakes} E.~L.~O.,  {Tielens} A.~G.~G.~M.,  1994, \apj, 427, 822

\bibitem[\protect\citeauthoryear{{Benincasa}, {Tasker}, {Pudritz} \&
  {Wadsley}}{{Benincasa} et~al.}{2013}]{2013ApJ...776...23B}
{Benincasa} S.~M.,  {Tasker} E.~J.,  {Pudritz} R.~E.,    {Wadsley} J.,  2013,
  \apj, 776, 23

\bibitem[\protect\citeauthoryear{{Berry}, {Reinhold}, {Jenness} \&
  {Economou}}{{Berry} et~al.}{2007}]{2007ASPC..376..425B}
{Berry} D.~S.,  {Reinhold} K.,  {Jenness} T.,    {Economou} F.,  2007, in
  {Shaw} R.~A.,  {Hill} F.,   {Bell} D.~J.,  eds, Astronomical Data Analysis
  Software and Systems XVI Vol.~376 of Astronomical Society of the Pacific
  Conference Series, {CUPID: A Clump Identification and Analysis Package}.
p.~425

\bibitem[\protect\citeauthoryear{{Bertoldi} \& {McKee}}{{Bertoldi} \&
  {McKee}}{1992}]{1992ApJ...395..140B}
{Bertoldi} F.,  {McKee} C.~F.,  1992, \apj, 395, 140

\bibitem[\protect\citeauthoryear{{Bigiel}, {Bolatto}, {Leroy}, {Blitz},
  {Walter}, {Rosolowsky}, {Lopez} \& {Plambeck}}{{Bigiel}
  et~al.}{2010}]{2010ApJ...725.1159B}
{Bigiel} F.,  {Bolatto} A.~D.,  {Leroy} A.~K.,  {Blitz} L.,  {Walter} F.,
  {Rosolowsky} E.~W.,  {Lopez} L.~A.,    {Plambeck} R.~L.,  2010, \apj, 725,
  1159

\bibitem[\protect\citeauthoryear{{Bolatto}, {Leroy}, {Rosolowsky}, {Walter} \&
  {Blitz}}{{Bolatto} et~al.}{2008}]{2008ApJ...686..948B}
{Bolatto} A.~D.,  {Leroy} A.~K.,  {Rosolowsky} E.,  {Walter} F.,    {Blitz} L.,
   2008, \apj, 686, 948

\bibitem[\protect\citeauthoryear{{Bolatto}, {Wolfire} \& {Leroy}}{{Bolatto}
  et~al.}{2013}]{2013ARA&A..51..207B}
{Bolatto} A.~D.,  {Wolfire} M.,    {Leroy} A.~K.,  2013, \araa, 51, 207

\bibitem[\protect\citeauthoryear{{Braun}, {Schmidt}, {Niemeyer} \&
  {Almgren}}{{Braun} et~al.}{2014}]{2014MNRAS.442.3407B}
{Braun} H.,  {Schmidt} W.,  {Niemeyer} J.~C.,    {Almgren} A.~S.,  2014,
  \mnras, 442, 3407

\bibitem[\protect\citeauthoryear{{Cald{\'u}-Primo}, {Schruba}, {Walter},
  {Leroy}, {Bolatto} \& {Vogel}}{{Cald{\'u}-Primo}
  et~al.}{2015}]{2015AJ....149...76C}
{Cald{\'u}-Primo} A.,  {Schruba} A.,  {Walter} F.,  {Leroy} A.,  {Bolatto}
  A.~D.,    {Vogel} S.,  2015, \aj, 149, 76

\bibitem[\protect\citeauthoryear{{Cen}}{{Cen}}{1992}]{1992ApJS...78..341C}
{Cen} R.,  1992, \apjs, 78, 341

\bibitem[\protect\citeauthoryear{{Colombo}, {Hughes}, {Schinnerer}, {Meidt},
  {Leroy}, {Pety}, {Dobbs}, {Garc{\'{\i}}a-Burillo}, {Dumas}, {Thompson},
  {Schuster} \& {Kramer}}{{Colombo} et~al.}{2014}]{2014ApJ...784....3C}
{Colombo} D.,  {Hughes} A.,  {Schinnerer} E.,  {Meidt} S.~E.,  {Leroy} A.~K.,
  {Pety} J.,  {Dobbs} C.~L.,  {Garc{\'{\i}}a-Burillo} S.,  {Dumas} G.,
  {Thompson} T.~A.,  {Schuster} K.~F.,    {Kramer} C.,  2014, \apj, 784, 3

\bibitem[\protect\citeauthoryear{{Combes}, {Garc{\'{\i}}a-Burillo}, {Casasola},
  {Hunt}, {Krips}, {Baker}, {Boone}, {Eckart}, {Marquez}, {Neri}, {Schinnerer}
  \& {Tacconi}}{{Combes} et~al.}{2014}]{2014A&A...565A..97C}
{Combes} F.,  {Garc{\'{\i}}a-Burillo} S.,  {Casasola} V.,  {Hunt} L.~K.,
  {Krips} M.,  {Baker} A.~J.,  {Boone} F.,  {Eckart} A.,  {Marquez} I.,  {Neri}
  R.,  {Schinnerer} E.,    {Tacconi} L.~J.,  2014, \aap, 565, A97

\bibitem[\protect\citeauthoryear{{Dame}, {Hartmann} \& {Thaddeus}}{{Dame}
  et~al.}{2001}]{2001ApJ...547..792D}
{Dame} T.~M.,  {Hartmann} D.,    {Thaddeus} P.,  2001, \apj, 547, 792

\bibitem[\protect\citeauthoryear{{Dickman}}{{Dickman}}{1975}]{1975ApJ...202...50D}
{Dickman} R.~L.,  1975, \apj, 202, 50

\bibitem[\protect\citeauthoryear{{Dickman}}{{Dickman}}{1978}]{1978ApJS...37..407D}
{Dickman} R.~L.,  1978, \apjs, 37, 407

\bibitem[\protect\citeauthoryear{{Dobbs}}{{Dobbs}}{2008}]{2008MNRAS.391..844D}
{Dobbs} C.~L.,  2008, \mnras, 391, 844

\bibitem[\protect\citeauthoryear{{Dobbs}, {Bonnell} \& {Pringle}}{{Dobbs}
  et~al.}{2006}]{2006MNRAS.371.1663D}
{Dobbs} C.~L.,  {Bonnell} I.~A.,    {Pringle} J.~E.,  2006, \mnras, 371, 1663

\bibitem[\protect\citeauthoryear{{Dobbs}, {Burkert} \& {Pringle}}{{Dobbs}
  et~al.}{2011}]{2011MNRAS.413.2935D}
{Dobbs} C.~L.,  {Burkert} A.,    {Pringle} J.~E.,  2011, \mnras, 413, 2935

\bibitem[\protect\citeauthoryear{{Dobbs}, {Glover}, {Clark} \&
  {Klessen}}{{Dobbs} et~al.}{2008}]{2008MNRAS.389.1097D}
{Dobbs} C.~L.,  {Glover} S.~C.~O.,  {Clark} P.~C.,    {Klessen} R.~S.,  2008,
  \mnras, 389, 1097

\bibitem[\protect\citeauthoryear{{Dobbs} \& {Pringle}}{{Dobbs} \&
  {Pringle}}{2013}]{2013MNRAS.432..653D}
{Dobbs} C.~L.,  {Pringle} J.~E.,  2013, \mnras, 432, 653

\bibitem[\protect\citeauthoryear{{Donovan Meyer}, {Koda}, {Momose}, {Mooney},
  {Egusa}, {Carty}, {Kennicutt}, {Kuno}, {Rebolledo}, {Sawada}, {Scoville} \&
  {Wong}}{{Donovan Meyer} et~al.}{2013}]{2013ApJ...772..107D}
{Donovan Meyer} J.,  {Koda} J.,  {Momose} R.,  {Mooney} T.,  {Egusa} F.,
  {Carty} M.,  {Kennicutt} R.,  {Kuno} N.,  {Rebolledo} D.,  {Sawada} T.,
  {Scoville} N.,    {Wong} T.,  2013, \apj, 772, 107

\bibitem[\protect\citeauthoryear{{Draine} \& {Bertoldi}}{{Draine} \&
  {Bertoldi}}{1996}]{1996ApJ...468..269D}
{Draine} B.~T.,  {Bertoldi} F.,  1996, \apj, 468, 269

\bibitem[\protect\citeauthoryear{{Engargiola}, {Plambeck}, {Rosolowsky} \&
  {Blitz}}{{Engargiola} et~al.}{2003}]{2003ApJS..149..343E}
{Engargiola} G.,  {Plambeck} R.~L.,  {Rosolowsky} E.,    {Blitz} L.,  2003,
  \apjs, 149, 343

\bibitem[\protect\citeauthoryear{{Feldmann}, {Gnedin} \& {Kravtsov}}{{Feldmann}
  et~al.}{2012}]{2012ApJ...758..127F}
{Feldmann} R.,  {Gnedin} N.~Y.,    {Kravtsov} A.~V.,  2012, \apj, 758, 127

\bibitem[\protect\citeauthoryear{{Fujimoto}, {Tasker}, {Wakayama} \&
  {Habe}}{{Fujimoto} et~al.}{2014}]{2014MNRAS.439..936F}
{Fujimoto} Y.,  {Tasker} E.~J.,  {Wakayama} M.,    {Habe} A.,  2014, \mnras,
  439, 936

\bibitem[\protect\citeauthoryear{{Galli} \& {Palla}}{{Galli} \&
  {Palla}}{1998}]{1998A&A...335..403G}
{Galli} D.,  {Palla} F.,  1998, \aap, 335, 403

\bibitem[\protect\citeauthoryear{{Glover} \& {Clark}}{{Glover} \&
  {Clark}}{2012}]{2012MNRAS.421..116G}
{Glover} S.~C.~O.,  {Clark} P.~C.,  2012, \mnras, 421, 116

\bibitem[\protect\citeauthoryear{{Glover}, {Federrath}, {Mac Low} \&
  {Klessen}}{{Glover} et~al.}{2010}]{2010MNRAS.404....2G}
{Glover} S.~C.~O.,  {Federrath} C.,  {Mac Low} M.-M.,    {Klessen} R.~S.,
  2010, \mnras, 404, 2

\bibitem[\protect\citeauthoryear{{Goldsmith} \& {Langer}}{{Goldsmith} \&
  {Langer}}{1978}]{1978ApJ...222..881G}
{Goldsmith} P.~F.,  {Langer} W.~D.,  1978, \apj, 222, 881

\bibitem[\protect\citeauthoryear{{Gratier}, {Braine}, {Rodriguez-Fernandez},
  {Schuster}, {Kramer}, {Corbelli}, {Combes}, {Brouillet}, {van der Werf} \&
  {R{\"o}llig}}{{Gratier} et~al.}{2012}]{2012A&A...542A.108G}
{Gratier} P.,  {Braine} J.,  {Rodriguez-Fernandez} N.~J.,  {Schuster} K.~F.,
  {Kramer} C.,  {Corbelli} E.,  {Combes} F.,  {Brouillet} N.,  {van der Werf}
  P.~P.,    {R{\"o}llig} M.,  2012, \aap, 542, A108

\bibitem[\protect\citeauthoryear{{Heyer}, {Krawczyk}, {Duval} \&
  {Jackson}}{{Heyer} et~al.}{2009}]{2009ApJ...699.1092H}
{Heyer} M.,  {Krawczyk} C.,  {Duval} J.,    {Jackson} J.~M.,  2009, \apj, 699,
  1092

\bibitem[\protect\citeauthoryear{Hindmarsh, Brown, Grant, Lee, Serban, Shumaker
  \& Woodward}{Hindmarsh et~al.}{2005}]{Hindmarsh2005}
Hindmarsh A.~C.,  Brown P.~N.,  Grant K.~E.,  Lee S.~L.,  Serban R.,  Shumaker
  D.~E.,    Woodward C.~S.,  2005, ACM Trans. Math. Softw., 31, 363

\bibitem[\protect\citeauthoryear{{Hollenbach} \& {McKee}}{{Hollenbach} \&
  {McKee}}{1979}]{1979ApJS...41..555H}
{Hollenbach} D.,  {McKee} C.~F.,  1979, \apjs, 41, 555

\bibitem[\protect\citeauthoryear{{Hollenbach} \& {McKee}}{{Hollenbach} \&
  {McKee}}{1989}]{1989ApJ...342..306H}
{Hollenbach} D.,  {McKee} C.~F.,  1989, \apj, 342, 306

\bibitem[\protect\citeauthoryear{{Hopkins}, {Quataert} \& {Murray}}{{Hopkins}
  et~al.}{2012}]{2012MNRAS.421.3488H}
{Hopkins} P.~F.,  {Quataert} E.,    {Murray} N.,  2012, \mnras, 421, 3488

\bibitem[\protect\citeauthoryear{{Hughes}, {Meidt}, {Colombo}, {Schinnerer},
  {Pety}, {Leroy}, {Dobbs}, {Garc{\'{\i}}a-Burillo}, {Thompson}, {Dumas},
  {Schuster} \& {Kramer}}{{Hughes} et~al.}{2013}]{2013ApJ...779...46H}
{Hughes} A.,  {Meidt} S.~E.,  {Colombo} D.,  {Schinnerer} E.,  {Pety} J.,
  {Leroy} A.~K.,  {Dobbs} C.~L.,  {Garc{\'{\i}}a-Burillo} S.,  {Thompson}
  T.~A.,  {Dumas} G.,  {Schuster} K.~F.,    {Kramer} C.,  2013, \apj, 779, 46

\bibitem[\protect\citeauthoryear{{Khoperskov}, {Zasov} \&
  {Tyurina}}{{Khoperskov} et~al.}{2003}]{2003ARep...47..357K}
{Khoperskov} A.~V.,  {Zasov} A.~V.,    {Tyurina} N.~V.,  2003, Astronomy
  Reports, 47, 357

\bibitem[\protect\citeauthoryear{{Khoperskov}, {Vasiliev}, {Khoperskov} \&
  {Lubimov}}{{Khoperskov} et~al.}{2014}]{2014JPhCS.510a2011K}
{Khoperskov} S.~A.,  {Vasiliev} E.~O.,  {Khoperskov} A.~V.,    {Lubimov} V.~N.,
   2014, Journal of Physics Conference Series, 510, 012011

\bibitem[\protect\citeauthoryear{{Khoperskov}, {Vasiliev}, {Sobolev} \&
  {Khoperskov}}{{Khoperskov} et~al.}{2013}]{2013MNRAS.428.2311K}
{Khoperskov} S.~A.,  {Vasiliev} E.~O.,  {Sobolev} A.~M.,    {Khoperskov} A.~V.,
   2013, \mnras, 428, 2311

\bibitem[\protect\citeauthoryear{{Koda} \& {et al.}}{{Koda} \& {et
  al.}}{2009}]{2009ApJ...700L.132K}
{Koda} {et al.} 2009, \apjl, 700, L132

\bibitem[\protect\citeauthoryear{{Kraljic}, {Renaud}, {Bournaud}, {Combes},
  {Elmegreen}, {Emsellem} \& {Teyssier}}{{Kraljic}
  et~al.}{2014}]{2014ApJ...784..112K}
{Kraljic} K.,  {Renaud} F.,  {Bournaud} F.,  {Combes} F.,  {Elmegreen} B.,
  {Emsellem} E.,    {Teyssier} R.,  2014, \apj, 784, 112

\bibitem[\protect\citeauthoryear{{Kritsuk}, {Lee} \& {Norman}}{{Kritsuk}
  et~al.}{2013}]{2013MNRAS.436.3247K}
{Kritsuk} A.~G.,  {Lee} C.~T.,    {Norman} M.~L.,  2013, \mnras, 436, 3247

\bibitem[\protect\citeauthoryear{{Larson}}{{Larson}}{1981}]{1981MNRAS.194..809L}
{Larson} R.~B.,  1981, \mnras, 194, 809

\bibitem[\protect\citeauthoryear{{Leitherer}, {Schaerer}, {Goldader},
  {Delgado}, {Robert}, {Kune}, {de Mello}, {Devost} \& {Heckman}}{{Leitherer}
  et~al.}{1999}]{sb99}
{Leitherer} C.,  {Schaerer} D.,  {Goldader} J.~D.,  {Delgado} R.~M.~G.,
  {Robert} C.,  {Kune} D.~F.,  {de Mello} D.~F.,  {Devost} D.,    {Heckman}
  T.~M.,  1999, \apjs, 123, 3

\bibitem[\protect\citeauthoryear{{Leroy}, {Walter}, {Bigiel}, {Usero}, {Weiss},
  {Brinks}, {de Blok}, {Kennicutt}, {Schuster}, {Kramer}, {Wiesemeyer} \&
  {Roussel}}{{Leroy} et~al.}{2009}]{2009AJ....137.4670L}
{Leroy} A.~K.,  {Walter} F.,  {Bigiel} F.,  {Usero} A.,  {Weiss} A.,  {Brinks}
  E.,  {de Blok} W.~J.~G.,  {Kennicutt} R.~C.,  {Schuster} K.-F.,  {Kramer} C.,
   {Wiesemeyer} H.~W.,    {Roussel} H.,  2009, \aj, 137, 4670

\bibitem[\protect\citeauthoryear{{Meidt}, {Hughes}, {Dobbs}, {Pety},
  {Thompson}, {Garcia-Burillo}, {Leroy}, {Schinnerer}, {Colombo}, {Querejeta},
  {Kramer}, {Schuster} \& {Dumas}}{{Meidt} et~al.}{2015}]{2015arXiv150404528M}
{Meidt} S.~E.,  {Hughes} A.,  {Dobbs} C.~L.,  {Pety} J.,  {Thompson} T.~A.,
  {Garcia-Burillo} S.,  {Leroy} A.~K.,  {Schinnerer} E.,  {Colombo} D.,
  {Querejeta} M.,  {Kramer} C.,  {Schuster} K.~F.,    {Dumas} G.,  2015, ArXiv
  e-prints

\bibitem[\protect\citeauthoryear{{Nelson} \& {Langer}}{{Nelson} \&
  {Langer}}{1999}]{1999ApJ...524..923N}
{Nelson} R.~P.,  {Langer} W.~D.,  1999, \apj, 524, 923

\bibitem[\protect\citeauthoryear{{Omukai}}{{Omukai}}{2000}]{2000ApJ...534..809O}
{Omukai} K.,  2000, \apj, 534, 809

\bibitem[\protect\citeauthoryear{{Renaud}, {Bournaud}, {Emsellem}, {Elmegreen},
  {Teyssier}, {Alves}, {Chapon}, {Combes}, {Dekel}, {Gabor}, {Hennebelle} \&
  {Kraljic}}{{Renaud} et~al.}{2013}]{2013MNRAS.436.1836R}
{Renaud} F.,  {Bournaud} F.,  {Emsellem} E.,  {Elmegreen} B.,  {Teyssier} R.,
  {Alves} J.,  {Chapon} D.,  {Combes} F.,  {Dekel} A.,  {Gabor} J.,
  {Hennebelle} P.,    {Kraljic} K.,  2013, \mnras, 436, 1836

\bibitem[\protect\citeauthoryear{{Roman-Duval}, {Jackson}, {Heyer}, {Johnson},
  {Rathborne}, {Shah} \& {Simon}}{{Roman-Duval}
  et~al.}{2009}]{2009ApJ...699.1153R}
{Roman-Duval} J.,  {Jackson} J.~M.,  {Heyer} M.,  {Johnson} A.,  {Rathborne}
  J.,  {Shah} R.,    {Simon} R.,  2009, \apj, 699, 1153

\bibitem[\protect\citeauthoryear{Roman-Duval, Jackson, Heyer, Rathborne \&
  Simon}{Roman-Duval et~al.}{2010}]{0004-637X-723-1-492}
Roman-Duval J.,  Jackson J.~M.,  Heyer M.,  Rathborne J.,    Simon R.,  2010,
  The Astrophysical Journal, 723, 492

\bibitem[\protect\citeauthoryear{{Roman-Duval}, {Jackson}, {Heyer}, {Rathborne}
  \& {Simon}}{{Roman-Duval} et~al.}{2010}]{2010ApJ...723..492R}
{Roman-Duval} J.,  {Jackson} J.~M.,  {Heyer} M.,  {Rathborne} J.,    {Simon}
  R.,  2010, \apj, 723, 492

\bibitem[\protect\citeauthoryear{{Romeo} \& {Wiegert}}{{Romeo} \&
  {Wiegert}}{2011}]{2011MNRAS.416.1191R}
{Romeo} A.~B.,  {Wiegert} J.,  2011, \mnras, 416, 1191

\bibitem[\protect\citeauthoryear{{Rosolowsky}}{{Rosolowsky}}{2007}]{2007ApJ...654..240R}
{Rosolowsky} E.,  2007, \apj, 654, 240

\bibitem[\protect\citeauthoryear{{Rosolowsky} \& {Leroy}}{{Rosolowsky} \&
  {Leroy}}{2006}]{2006PASP..118..590R}
{Rosolowsky} E.,  {Leroy} A.,  2006, \pasp, 118, 590

\bibitem[\protect\citeauthoryear{{Schinnerer}, {Meidt}, {Pety}, {Hughes},
  {Colombo}, {Garc{\'{\i}}a-Burillo}, {Schuster}, {Dumas}, {Dobbs}, {Leroy},
  {Kramer}, {Thompson} \& {Regan}}{{Schinnerer}
  et~al.}{2013}]{2013ApJ...779...42S}
{Schinnerer} E.,  {Meidt} S.~E.,  {Pety} J.,  {Hughes} A.,  {Colombo} D.,
  {Garc{\'{\i}}a-Burillo} S.,  {Schuster} K.~F.,  {Dumas} G.,  {Dobbs} C.~L.,
  {Leroy} A.~K.,  {Kramer} C.,  {Thompson} T.~A.,    {Regan} M.~W.,  2013,
  \apj, 779, 42

\bibitem[\protect\citeauthoryear{{Scoville}, {Solomon} \& {Sanders}}{{Scoville}
  et~al.}{1979}]{1979IAUS...84..277S}
{Scoville} N.~Z.,  {Solomon} P.~M.,    {Sanders} D.~B.,  1979, in {Burton}
  W.~B.,  ed., The Large-Scale Characteristics of the Galaxy Vol.~84 of IAU
  Symposium, {CO observations of spiral structure and the lifetime of giant
  molecular clouds}.
pp 277--282

\bibitem[\protect\citeauthoryear{{Shetty} \& {Ostriker}}{{Shetty} \&
  {Ostriker}}{2008}]{2008ApJ...684..978S}
{Shetty} R.,  {Ostriker} E.~C.,  2008, \apj, 684, 978

\bibitem[\protect\citeauthoryear{{Solomon}, {Rivolo}, {Barrett} \&
  {Yahil}}{{Solomon} et~al.}{1987}]{1987ApJ...319..730S}
{Solomon} P.~M.,  {Rivolo} A.~R.,  {Barrett} J.,    {Yahil} A.,  1987, \apj,
  319, 730

\bibitem[\protect\citeauthoryear{{Tan}, {Leech}, {Rigopoulou} \& et.al.}{{Tan}
  et~al.}{2013}]{2013MNRAS.436..921T}
{Tan} B.-K.,  {Leech} J.,  {Rigopoulou} D.,    et.al. 2013, \mnras, 436, 921

\bibitem[\protect\citeauthoryear{{Tasker}}{{Tasker}}{2011}]{2011ApJ...730...11T}
{Tasker} E.~J.,  2011, \apj, 730, 11

\bibitem[\protect\citeauthoryear{{Tasker} \& {Tan}}{{Tasker} \&
  {Tan}}{2009}]{2009ApJ...700..358T}
{Tasker} E.~J.,  {Tan} J.~C.,  2009, \apj, 700, 358

\bibitem[\protect\citeauthoryear{{Visser}, {van Dishoeck} \& {Black}}{{Visser}
  et~al.}{2009}]{2009A&A...503..323V}
{Visser} R.,  {van Dishoeck} E.~F.,    {Black} J.~H.,  2009, \aap, 503, 323

\bibitem[\protect\citeauthoryear{{Williams}, {de Geus} \& {Blitz}}{{Williams}
  et~al.}{1994}]{1994ApJ...428..693W}
{Williams} J.~P.,  {de Geus} E.~J.,    {Blitz} L.,  1994, \apj, 428, 693

\bibitem[\protect\citeauthoryear{{Williams} \& {McKee}}{{Williams} \&
  {McKee}}{1997}]{1997ApJ...476..166W}
{Williams} J.~P.,  {McKee} C.~F.,  1997, \apj, 476, 166

\bibitem[\protect\citeauthoryear{{Wolfire}, {McKee}, {Hollenbach} \&
  {Tielens}}{{Wolfire} et~al.}{2003}]{2003ApJ...587..278W}
{Wolfire} M.~G.,  {McKee} C.~F.,  {Hollenbach} D.,    {Tielens} A.~G.~G.~M.,
  2003, \apj, 587, 278

\bibitem[\protect\citeauthoryear{{Zasov} \& {Kasparova}}{{Zasov} \&
  {Kasparova}}{2014}]{2014Ap&SS.353..595Z}
{Zasov} A.,  {Kasparova} A.,  2014, \apss, 353, 595

\end{thebibliography}

\end{document}